\documentclass[longauth]{aa}

\usepackage[varg]{txfonts}
\usepackage{natbib}
\usepackage{multirow}
\usepackage{graphicx}   
\usepackage[x11names]{xcolor}

\usepackage{hyperref}
\hypersetup{
      colorlinks = false,
      linkcolor = .,
      linkbordercolor = green,
}

\newcommand{\rband}{{$r$-band}}

\newcommand{\utok}{\ensuremath{ugri_{1}i_{2}ZY\!J\!H\!K_{\rm s}}}

\newcommand{\kidz}{{KiDZ}}
\newcommand{\kids}{{KiDS}}

\newcommand{\konek}{{KiDS-$1000$}}
\newcommand{\kidslegacy}{{KiDS-Legacy}}
\newcommand{\skills}{{SKiLLS}}

\newcommand{\viking}{{VIKING}}
\newcommand{\euclid}{{\it Euclid}}

\newcommand{\gaap}{{\sc GAaP}}

\newcommand{\cosmopipe}{{\sc CosmoPipe}}

\newcommand{\lensfit}{{\textit{lens}fit}}
\newcommand{\sourceextractor}{{\sc Source Extractor}}

\newcommand{\yaw}{\texttt{yet\_another\_wizz}}

\newcommand{\drfive}{{\sc dr5}}
\newcommand{\photoz}{{photo\nobreakdash-$z$}}
\newcommand{\specz}{{spec\nobreakdash-$z$}}
\newcommand{\zb}{\ensuremath{z_{\rm B}}}

\newcommand{\nz}{\ensuremath{N(z)}}
\newcommand{\truenz}{\ensuremath{N_{\rm true}(z)}}
\newcommand{\somnz}{\ensuremath{N_{\rm SOM}(z)}}
\newcommand{\ccnz}{\ensuremath{N_{\rm CC}(z)}}

\newcommand{\sqdeg}{\ensuremath{\text{deg}^2}}
\newcommand{\perarcminsq}{{\ensuremath{\text{arcmin}^{-2}}}}
\newcommand{\kpc}{{\ensuremath{\text{kpc}}}}

\newcommand{\nolinenumbers}{{}}

\newcommand{\dshift}[1]{\ensuremath{D_{\rm z}^{\rm #1}}}

\newcommand\Tstrut{\rule{0pt}{2.4ex}}
\newcommand\Bstrut{\rule[-1.0ex]{0pt}{0pt}}

\begin{document}

\defcitealias{kuijken/etal:2019}{DR4}
\defcitealias{wright/etal:2024}{W24}

\title{\kidslegacy: Redshift distributions and their calibration}
\titlerunning{\kidslegacy\ \nz\ Calibration}

\author{ Angus~H.~Wright                \inst{1} \thanks{awright@astro.rub.de} \and 
Hendrik~Hildebrandt              \inst{1} \and   
Jan~Luca~van~den~Busch            \inst{1} \and    
Maciej~Bilicki                    \inst{2} \and    
Catherine~Heymans                \inst{1,3} \and  
Benjamin~Joachimi                \inst{4} \and    
Constance~Mahony                        \inst{1,5,6} \and 
Robert~Reischke                  \inst{1,7} \and 
Benjamin~St\"olzner              \inst{1} \and  
Anna~Wittje                          \inst{1} \and  
Marika~Asgari                    \inst{8} \and  
Nora~Elisa~Chisari               \inst{9,10} \and  
Andrej~Dvornik                    \inst{1} \and   
Christos~Georgiou                \inst{11} \and   
Benjamin~Giblin                  \inst{3} \and     
Henk~Hoekstra                    \inst{10} \and      
Priyanka~Jalan                   \inst{2} \and       
Anjitha~John~William             \inst{2} \and
Shahab~Joudaki                   \inst{12,13} \and   
Konrad~Kuijken                   \inst{10} \and      
Giorgio~Francesco~Lesci          \inst{14,15} \and   
Shun-Sheng~Li                    \inst{1,10} \and    
Laila~Linke                      \inst{16} \and      
Arthur~Loureiro                  \inst{17,18} \and   
Matteo~Maturi                    \inst{19} \and      
Lauro~Moscardini                 \inst{14,15,20} \and
Lucas~Porth                      \inst{7} \and      
Mario~Radovich                   \inst{21} \and     
Tilman~Tr\"oster                 \inst{22} \and     
Maximilian~von~Wietersheim-Kramsta \inst{23,24} \and
Ziang~Yan                        \inst{1} \and
Mijin~Yoon                       \inst{10} \and
Yun-Hao~Zhang \inst{1}           \inst{3,10}            
}
\authorrunning{The KiDS Collaboration}
\institute{
Ruhr University Bochum, Faculty of Physics and Astronomy, Astronomical Institute (AIRUB), German Centre for Cosmological Lensing, 44780 Bochum, Germany \and
Center for Theoretical Physics, Polish Academy of Sciences, al. Lotników 32/46, 02-668 Warsaw, Poland \and
Institute for Astronomy, University of Edinburgh, Royal Observatory, Blackford Hill, Edinburgh, EH9 3HJ, UK. \and
Department of Physics and Astronomy, University College London, Gower Street, London WC1E 6BT, UK \and
Department of Physics, University of Oxford, Denys Wilkinson Building, Keble Road, Oxford OX1 3RH, United Kingdom. \and
Donostia International Physics Center, Manuel Lardizabal Ibilbidea, 4, 20018 Donostia, Gipuzkoa, Spain. \and
Argelander-Institut für Astronomie, Universität Bonn, Auf dem Hügel 71, D-53121 Bonn, Germany \and
School of Mathematics, Statistics and Physics, Newcastle University, Herschel Building, NE1 7RU, Newcastle-upon-Tyne, UK \and
Institute for Theoretical Physics, Utrecht University, Princetonplein 5, 3584CC Utrecht, The Netherlands. \and
Leiden Observatory, Leiden University, P.O.Box 9513, 2300RA Leiden, The Netherlands \and
Institut de Física d’Altes Energies (IFAE), The Barcelona Institute of Science and Technology, Campus UAB, 08193 Bellaterra (Barcelona), Spain \and
Centro de Investigaciones Energéticas, Medioambientales y Tecnológicas (CIEMAT), Av. Complutense 40, E-28040 Madrid, Spain \and
Institute of Cosmology \& Gravitation, Dennis Sciama Building, University of Portsmouth, Portsmouth, PO1 3FX, United Kingdom \and
Dipartimento di Fisica e Astronomia "Augusto Righi" - Alma Mater Studiorum Università di Bologna, via Piero Gobetti 93/2, 40129 Bologna, Italy \and
INAF-Osservatorio di Astrofisica e Scienza dello Spazio di Bologna, Via Piero Gobetti 93/3, 40129 Bologna, Italy \and
Universität Innsbruck, Institut für Astro- und Teilchenphysik, Technikerstr. 25/8, 6020 Innsbruck, Austria \and
The Oskar Klein Centre, Department of Physics, Stockholm University, AlbaNova University Centre, SE-106 91 Stockholm, Sweden \and
Imperial Centre for Inference and Cosmology (ICIC), Blackett Laboratory, Imperial College London, Prince Consort Road, London SW7 2AZ, UK \and
Zentrum für Astronomie, Universitatät Heidelberg, Philosophenweg 12, D-69120 Heidelberg, Germany; Institute for Theoretical Physics, Philosophenweg 16, D-69120 Heidelberg, Germany \and
Istituto Nazionale di Fisica Nucleare (INFN) - Sezione di Bologna, viale Berti Pichat 6/2, I-40127 Bologna, Italy \and
INAF - Osservatorio Astronomico di Padova, via dell'Osservatorio 5, 35122 Padova, Italy \and
Institute for Particle Physics and Astrophysics, ETH Zürich, Wolfgang-Pauli-Strasse 27, 8093 Zürich, Switzerland \and
Institute for Computational Cosmology, Ogden Centre for Fundament Physics - West, Department of Physics, Durham University, South Road, Durham DH1 3LE, UK. \and
Centre for Extragalactic Astronomy, Ogden Centre for Fundament Physics - West, Department of Physics, Durham University, South Road, Durham DH1 3LE, UK 
}

\date{Received 31 March 2025 / Accepted 26 August 2025 }

\abstract 
{ 
We present the redshift calibration methodology and bias estimates for the cosmic shear analysis of 
the fifth and final data release (DR5) of the Kilo-Degree Survey (KiDS). KiDS-DR5 includes a greatly expanded 
compilation of calibrating spectra, drawn from $27$ square degrees of dedicated optical and near-IR imaging taken over 
deep spectroscopic fields. The redshift distribution calibration leverages a range of 
new methods and updated simulations to produce the most precise \nz\ bias estimates used by KiDS to date. 
Improvements to our colour-based redshift distribution measurement method using self-organising maps (SOMs) mean that we
are able to use many more sources per tomographic bin for our cosmological analyses and better estimate the
representation of our source sample given the available spec-$z$. We validated our colour-based redshift distribution
estimates with spectroscopic cross-correlations (CCs). We find that improvements to our CC redshift
distribution measurement methods mean that redshift distribution biases estimated between the SOM and CC methods are
fully consistent on simulations, and the data calibration is consistent to better than $2\sigma$ in all tomographic
bins. 
}


\maketitle
\tableofcontents

\section{Introduction}
\label{sec:intro}

Wide-field imaging surveys with large mosaic CCD cameras and
broadband optical and near-infrared (NIR) filters have entered a
crucial era where significant fractions of the sky are currently being
surveyed. The current generation called stage-III
\citep{sevilla/etal:2021,aihara/etal:2022,wright/etal:2024} covers
areas of more than a thousand square degrees and
will soon be superseded by stage-IV surveys
\citep{mellier/etal:2024,ivezic/etal:2019} covering an order of
magnitude larger areas at similar or greater depths. Perhaps the most
crucial analysis step for virtually any application of these surveys
is to add information about the radial distance of the very large
number of objects (typically of the order of $10^7$ -- $10^9$) reliably detected
in such surveys. In the absence of spectroscopic redshifts for these
huge samples of (mostly) galaxies, photometric redshifts
\citep[photo-$z$; for a recent review see][]{newman/etal:2022} based
on broadband multi-colour photometry are used to solve this problem.

The estimation of these broadband photo-$z$ for faint targets has
been surprisingly stable over the past two decades \citep{hildebrandt/etal:2010}. All stage-III
surveys base their main scientific analyses still on template-fitting
techniques developed more than 20 years ago \citep[e.g.][]{benitez:2000}. This reflects the
maturity of these techniques and their close-to optimal use of
information. Until the arrival of large, complete spectroscopic
training sets down to the magnitude limits of these wide-field imaging
surveys, which would enable highly precise and accurate photometric redshifts estimated 
via machine-learning techniques, this situation is unlikely to change \citep{newman/etal:2015}.

These photo-$z$ estimates of individual galaxies have well
characterised error distributions with typical scatter of a few per
cent around the true redshifts and equally a fraction of a few per
cent of catastrophic outliers. These numbers have
essentially remained unchanged for a long time. The main reason for this
perceived stagnation in individual photo-$z$ quality is the fact that
this performance is not the limiting factor for the main science driver of
such imaging surveys: weak gravitational lensing (WL).

The gravitational lensing effect is integrated along the line of sight
and -- in the case of WL -- measured statistically by averaging over
shear estimates of very large ensembles of galaxies. As such, a
significant improvement in individual galaxy photo-$z$ is not required. 
Instead, individual galaxy photo-$z$ values
are used only to divide the galaxy distribution into relatively broad,
so-called tomographic bins (hundreds of Mpc comoving) along the line of sight.

It is the ensemble redshift distribution, $\nz$, that has rightly received most attention in WL measurements taken in
the recent past as its accuracy is directly related to that of the cosmological parameters estimated from WL surveys.
The increasing statistical power, hence, comes with a paralleled increase in the required accuracy of these $\nz$, most
importantly expressed by their mean redshifts \citep{huterer/etal:2006}. Higher-order moments of the $\nz$ are less
important for cosmic shear but very relevant for other probes like galaxy clustering
\cite{mcloud/etal:2017,reischke:2024}. Here, we concentrate on the former and leave the quantification of calibration
uncertainties on higher-order moments of the $\nz$, such as width and skewness, to future work.  For the
current-generation stage-III surveys, the mean redshifts must be controlled at the percent level
\citep{myles/etal:2021,rau/etal:2023,hildebrandt/etal:2021}.  Any larger bias in the redshifts would lead to a bias in
the cosmological conclusions that would rival the statistical uncertainty. Calibration techniques are used to estimate
the $\nz$, and simulations are employed to estimate residual biases, which can be used to re-calibrate the data. The
uncertainty in this re-calibration is typically marginalised over in the cosmological inference.

The Kilo-Degree Survey \citep[KiDS;][]{wright/etal:2024} is conducted with OmegaCam mounted at the Cassegrain focus of
the European Southern Observatory VLT Survey Telescope (VST) on Paranal, Chile and complemented by the
VISTA\footnote{Visible and Infrared Survey Telescope for Astronomy} Kilo Degree Infrared Galaxy Survey
\citep[VIKING;][]{edge/etal:2013} observed from a neighbouring mountaintop. Together, these two surveys form a unique
nine-band dataset covering the near-UV to NIR with (in terms of depth) well-matched high-resolution images over an area
of $\sim1\,350$~\sqdeg. With this extensive filter coverage, KiDS has the potential of estimating well-controlled
photo-$z$ down to its magnitude limit ($r\sim24$ for a typical WL source at a S/N of $\sim10$) and estimating accurate
$\nz$ all the way to $z\la2$, paving the way for similar multi-camera, optical+NIR efforts with, for example, \euclid.

While individual galaxy photo-$z$ and their quality for the complete KiDS dataset are covered in the Data Release 5
(DR5) paper \citep[][]{wright/etal:2024}, here we describe the redshift calibration approach; that is, the estimation of the
$\nz$, and their characterisation with simulations. This is the final paper in a list of publications that have
developed the KiDS redshift calibration strategy
\citep{hildebrandt/etal:2017,hildebrandt/etal:2020,hildebrandt/etal:2021,wright/etal:2020a,vandenbusch/etal:2020,vandenbusch/etal:2022}.
Similar to previous efforts, we used two complementary techniques to estimate the $\nz$: one that is colour-based and
another that is position-based. Both of these techniques leverage the power of spectroscopic surveys that overlap with
KiDS or the newly compiled KiDZ dataset (i.e. the KiDS redshift calibration fields). The kinds of spectroscopic surveys
used for the two techniques are quite different, though, which is highly beneficial for systematic robustness and
independence of these methods.

The KiDS data, the KiDZ calibration fields, the calibrating spectroscopic surveys, and the tomographic binning approach
are described in Sect.~\ref{sec:data}. The mock catalogues that mimic these different datasets are introduced in
Sect.~\ref{sec:simulations}. In Sect.~\ref{sec:SOM}, the colour-based calibration technique via a self-organising map
(SOM) projection of the 9D colour space is introduced. This is complemented by a description of the position-based
calibration technique, also known as clustering redshifts (or dubbed CC for cross-correlation), in Sect.~\ref{sec:CC}.
The performance of these two approaches was evaluated on the simulated mock catalogues and is presented in
Sect.~\ref{results: simulations}. Results on the KiDS and KiDZ data are shown in Sect.~\ref{results: data}, which are
further discussed in Sect.~\ref{sec:discussion} before we summarise in Sect.~\ref{sec:summary}.
 
\section{Data}\label{sec:data}

This manuscript presents estimates of redshift distributions for the wide-field galaxy samples used in \kidslegacy. The
\kidslegacy\ dataset is described at length in the \kids\ \drfive\ data release document \citep[][hereafter
W24]{wright/etal:2024}. Here we summarise the pertinent information from the release including references to precise
sections therein. We direct the interested reader to the data release document for detailed information regarding the
data. 

\begin{table}
  \caption{Summary of relevant imaging data released in \kids-\drfive\ (including \kidz\ data).}
  \label{tab:summary}
  \centering
  \resizebox{\columnwidth}{!}{
  \begin{tabular}{ccrcc}
    \hline\hline
    Telescope \& & Filter & \multicolumn{1}{c}{$\lambda_{\rm cen}$} & Mag. Lim. & PSF FWHM \Tstrut \\
    Camera & & \multicolumn{1}{c}{$(\AA)$} & ($5\sigma\,2^{\prime\prime}$ AB) & ($^{\prime\prime}$) \\
    \hline\Tstrut
    \multirow{5}{20mm}{\centering VST (OmegaCAM)}
              & $u$         & $3\,550 $ & $24.26\pm0.10$& $1.01 \pm 0.17$ \\
              & $g$         & $4\,775 $ & $25.15\pm0.12$& $0.88 \pm 0.15$ \\
              & $r$         & $6\,230 $ & $25.07\pm0.14$& $0.70 \pm 0.12$ \\
              & $i_1$       & $7\,630 $ & $23.66\pm0.25$& $0.81 \pm 0.18$ \\
              & $i_2$       & $7\,630 $ & $23.73\pm0.30$& $0.81 \pm 0.18$ \\
    \hline\Tstrut
    \multirow{5}{20mm}{\centering VISTA (VIRCAM)}
             & $Z$         & $8\,770 $ & $23.79\pm0.20$ & $0.90\pm0.10$ \\
             & $Y$         & $10\,200$ & $23.02\pm0.19$ & $0.86\pm0.09$ \\
             & $J$         & $12\,520$ & $22.72\pm0.20$ & $0.85\pm0.07$ \\
             & $H$         & $16\,450$ & $22.27\pm0.24$ & $0.88\pm0.09$ \\
             & $K_{\rm s}$ & $21\,470$ & $22.02\pm0.19$ & $0.87\pm0.08$ \\
    \hline
  \end{tabular}
  }
\end{table}

The fifth data release of \kids\ consists of $1347$~\sqdeg\ of weak lensing imaging data, and $27$~\sqdeg\ of imaging
covering deep spectroscopic calibration fields (with $4$~\sqdeg\ of overlap). All data were observed with both VST and VISTA, yielding photometry in nine distinct
photometric bandpasses (four optical and five NIR). Additionally, the entire wide and calibration footprint was
observed twice in the $i$ band, yielding two realisations and epochs of the photometry in this band. These realisations are kept
separate in our analysis and are labelled $i_1$ and $i_2$ for distinction (the impact of the additional $i$-band measurements on 
our photo-$z$ is shown in \citetalias{wright/etal:2024}). This leads to a final dataset containing ten
photometric bands, which are summarised in Table~\ref{tab:summary}. Sources in these fields were extracted from the VST $r$-band
imaging using \sourceextractor\ \citep{bertin/arnouts:1996}, within the Astro-WISE analysis environment
\citep{valentijn/etal:2007,begeman/etal:2013,macfarland/etal:2013}, yeilding approximately $139$ million unique sources. 

All sources in \kids-\drfive\ have photometric information measured in all available photometric bands. This photometric
information was estimated through a form of matched aperture photometry that ensures consistent flux information is extracted
from each source across the ten photometric bands, based on the optical \rband, and accounting for variations in the point
spread function (PSF) per band. This forced photometry was performed with the Gaussian aperture and PSF code
(GAaP; \citealt{kuijken:2008}), and details of the implementation of GAaP in the context of \kids-\drfive\ can be found
in Sections
3.6 and 6 of \citetalias{wright/etal:2024}. 

After measurement of photometric information in all bands, the
\kids-\drfive\ sample was masked to include only unique sources that reside in
areas of high-quality data in all bands. This masking process is
described at length in Section~6.4 of \citetalias{wright/etal:2024}
and results in $100\,744\,685$ sources drawn from an effective area of
$1014.013$~\sqdeg (corresponding to an effective number density of
$10.94~\perarcminsq$). 

The lensing portion of the \kids-\drfive\ sample was given the name \kidslegacy. As in previous \kids\ analyses, the lensing 
sample contains per-source shape measurements and corresponding shape-measurement confidence weights estimated using the
\lensfit\ algorithm \citep{miller/etal:2007,miller/etal:2013}. 
These shapes were then calibrated with complex image simulations designed to emulate the properties of the
\kidslegacy\ sample as closely as possible. A detailed description of these simulations is given in \cite{li/etal:2022}, and
they are also summarised here in Sect.~\ref{sec:simulations}. The definition of the sample is provided in detail in Section~7.2 of
\citetalias{wright/etal:2024}, and involves a series of cuts in magnitude, colour, neighbour distance on-sky, and 
shape-measurement quality metrics. Additionally, \citet{wright/etal:2025b} found that masking of areas with higher
astrometric noise was required to satisfy their cosmic shear B-mode null tests, leading to an additional masking of the
survey footprint. The final \kidslegacy\ lensing sample is defined as the remaining $40\,950\,607$ sources after these
selections, drawn from $967.4$~\sqdeg\ (corresponding to an effective number density of $8.81~\perarcminsq$).

\subsection{Calibration datasets}\label{sec:calib_data}

\begin{figure}
  \centering
  \includegraphics[width=\columnwidth]{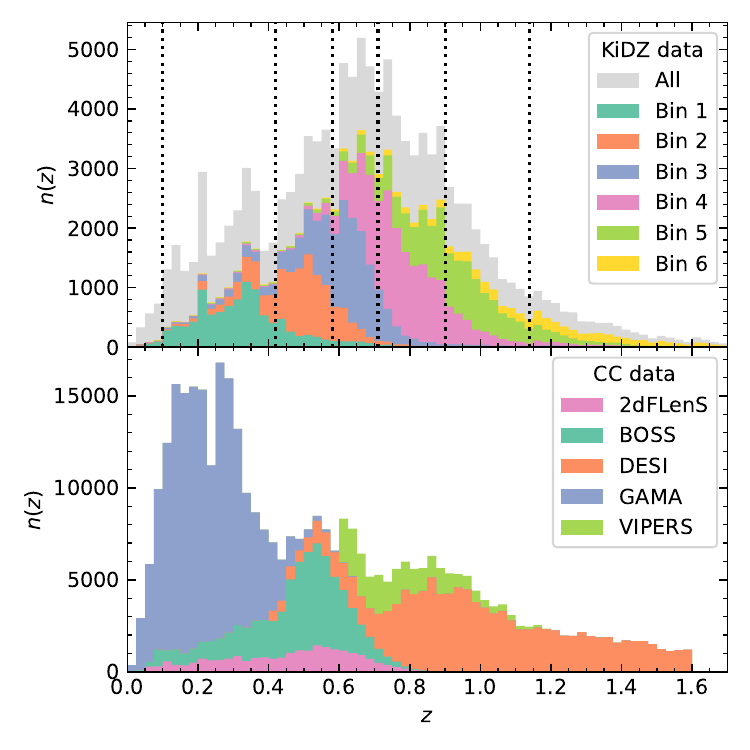}
  \caption{
    Redshift distribution of the calibration data used for \kidslegacy. The top panel displays the full KiDZ data in grey and the
    proportion of it that enters each tomographic bin after calibrating the fiducial SOM, shown as a stacked histogram. The bin edges are
    indicated by the vertical dashed lines. The bottom panel shows the spectroscopic surveys used as calibration samples for the
    clustering redshift measurements (also stacked).
  }
  \label{fig:spec_nz}
\end{figure}

The calibration sample used to estimate redshift distributions in \kidslegacy\ with the colour-based SOM method is drawn
principally from the \kidz\ sample described in Section~5 of \citetalias{wright/etal:2024}. The \kidz\ imaging were
taken under the same observational conditions as the main \kids\ and \viking\ surveys and thus share the photometric
properties of the wide-field dataset.  The sample consists of $126\,085$ sources drawn from $22$ spectroscopic samples
and/or surveys, which have been compiled following a hierarchy that resolves internal and external duplicates in the
datasets. The hierarchy ranks the constituents such that we kept spectra preferentially from the sample that is most
likely to provide a reliable redshift. The details of this hierarchy and the homogenisation of the various redshift
quality metrics are detailed in \citetalias{wright/etal:2024}, and the resulting redshift distribution is shown in the
top panel of Fig~\ref{fig:spec_nz}.

\begin{figure*}
  \centering
  \includegraphics[width=\textwidth]{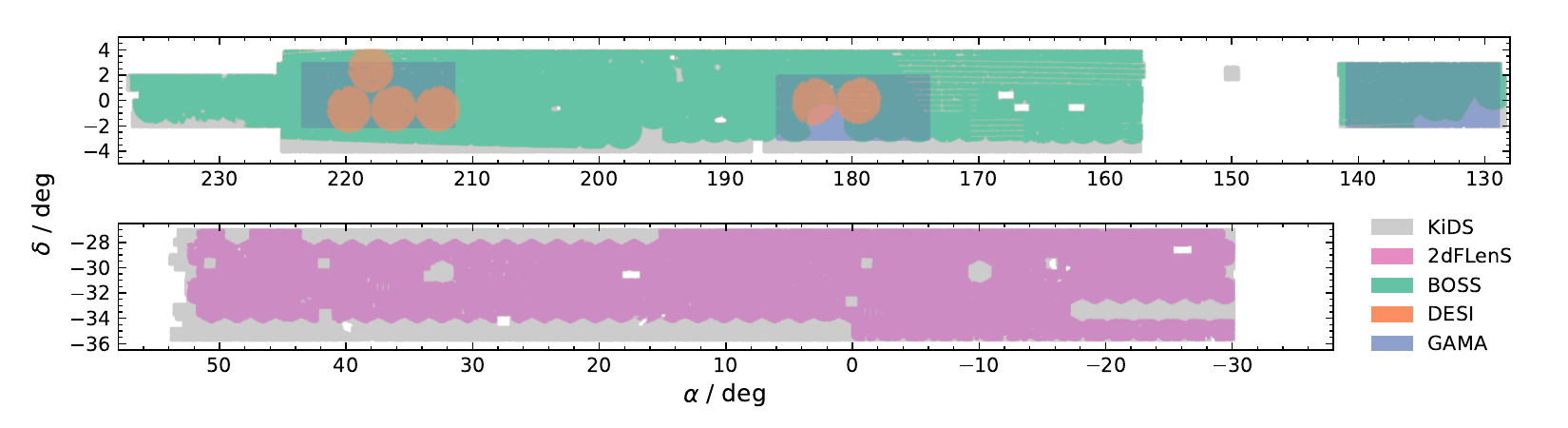}
  \caption{
    Footprint of the spectroscopic surveys overlapping \kids\ used for the clustering redshift measurements. VIPERS is exclusive
    to the \kidz\ fields, which are not shown here.
  }
  \label{fig:CZ_specz_footprint}
\end{figure*}

The calibration sample for clustering redshifts used in \kidslegacy\ differs from the one described in
\citetalias{wright/etal:2024}.  As opposed to previous work \citep{vandenbusch/etal:2020,hildebrandt/etal:2021}, we only
included samples that cover multiple KiDS tiles and provide sufficient contiguous overlap with \kids\ or \kidz\
observations, namely: the 2-degree Field Lensing Survey \citep[2dFLenS,][]{blake/etal:2016}, Sloan Digital Sky Survey
(SDSS, DR12) Baryon Osscilation Spectroscopic Survey (BOSS) low redshift (LOWZ) and constant mass (CMASS)
samples \citep{alam/etal:2015}, Galaxy and Mass Assembly (GAMA) DR4 \citep{driver/etal:2022}, and the VIMOS Public
Extragalactic Redshift Survey (VIPERS) \mbox{PDR-2} \citep{scodeggio/etal:2018}. We applied additional masking to ensure
a consistent footprint between the \kidslegacy\ data, the spectroscopic data, and their provided spectroscopic random
catalogues. We removed the relatively small overlap of 2dFLenS with the northern \kids\ patch and limited the VIPERS
dataset to a redshift range of $0.6 \leq z < 1.18$ to be consistent with the random catalogues and to mitigate the
incompleteness from the colour sampling at $z < 0.6$ \citep{garilli/etal:2014}. Finally, we added $109\,381$ recently
released spectra from the Dark Energy Spectroscopic Instrument \citep[DESI,][]{desi/etal:2016a,desi/etal:2016b} Early Data
Release. Specifically, we used the designated clustering catalogues containing a subset of the luminous red galaxy (LRG)
and emission line galaxy (ELG) samples \citep[see Section~4.2 of][]{desi/etal:2023}. This new set of calibration samples
for the cross-correlation (CC) method (bottom panel of Fig.~\ref{fig:spec_nz}) covers a combined total of more than
80~\% of the \kidslegacy\ footprint (Fig.~\ref{fig:CZ_specz_footprint}).

Table~\ref{tab: specz} details all datasets utilised for calibration in \kidslegacy. The table indicates samples that
are used for our colour-based SOM redshift calibration (Sect.~\ref{sec:SOM}) and for our position-based clustering redshift 
calibration (Sect.~\ref{sec:CC}). Their respective redshift distributions are shown in Fig.~\ref{fig:spec_nz}.

\begin{table}
  \centering
  \caption{Spectroscopic redshift samples used for the \kidslegacy\ redshift calibration.}\label{tab: specz}
  \resizebox{\columnwidth}{!}{
  \begin{tabular}{lrrcl}
    \hline
    \hline
    \multicolumn{1}{c}{Survey or Field} & \multicolumn{1}{c}{$N_{\rm spec}$}
    & \multicolumn{1}{c}{Area} & \multicolumn{1}{c}{Density} & \multicolumn{1}{c}{Usage} \Tstrut \\
     & & \multicolumn{1}{c}{[\sqdeg]} & \multicolumn{1}{c}{[\perarcminsq]} & \\
    \hline\Tstrut
    \kidz\ compilation & $126\,085$ & $ 19.3$ & $3.77$ & SOM \\
    \hline\Tstrut
    2dFLenS            & $ 22\,675$ & $382.4$ & $0.02$ & CC \\
    BOSS DR12          & $ 60\,482$ & $422.6$ & $0.04$ & CC \\
    DESI EDR           & $109\,381$ & $ 44.2$ & $0.69$ & CC \\
    GAMA DR4           & $161\,839$ & $136.1$ & $0.33$ & CC \\
    VIPERS             & $ 26\,408$ & $  9.3$ & $0.79$ & CC \\
    \hline
  \end{tabular}
  }
  \tablefoot{The \kidz\ spectroscopic compilation is described in \citetalias{wright/etal:2024}. VIPERS data are included in both the \kidz\ compilation used by the SOM and in the sample used for CCs.}
\end{table}

\subsection{Weight assignment}

One important distinction between the calibration fields and the wide-fields used for lensing is that the calibration fields
lack the data-products required for \lensfit\ shape estimation \citepalias[specifically individual calibrated exposures; see][]{wright/etal:2024}. As such, sources in calibration fields that do not overlap
with the wide-field data do not contain shape-measurement information, in particular the shape-measurement
weights (see \citetalias{wright/etal:2024} for details about the imaging differences within the \kidz\ fields).
Since the shape weights correlate with the photometric observables, 
they present an additional selection that has to be taken into account in the SOM and the CC calibration. 

Therefore, we replicated the \lensfit\ weights in \kidz\ using $k$-nearest neighbour matching. To each \kidz\ galaxy we assigned the
\lensfit\ weight of a galaxy from \kids-\drfive\ that is closest in \rband\ magnitude (\texttt{MAG\_AUTO}), half-light radius
(\texttt{FLUX\_RADIUS}), \gaap\ major-to-minor axis ratio (\texttt{Bgaper}~/ \texttt{Agaper}), photometric redshift (\texttt{Z\_B}),
and average PSF size per tile (\texttt{PSF\_RAD}). This matching process is therefore conditional on the \kids\ and
\kidz\ data having consistent photometric properties, in particular in the $r$ band (from which all but the photometric
redshift metric are defined). Fortunately, the observation of the \kidz\ data was taken with the same
observational requirements as the main survey data, and as such the imaging are very well matched: for example, the
distribution of \kidz\ $r$-band seeing and magnitude limits (which drive the statistical properties of measured sizes and
fluxes) are consistent with having been drawn randomly from the parent population of \kids\ $r$-band seeing and
magnitude limits, as computed using the two-sample \citet{anderson/darling:1954} test ($p$-values of 0.66 and 0.23,
respectively). 

To validate the accuracy of the matching process itself, we split the \kidslegacy\ wide-field sample into two halves
(by splitting the survey at RA$=180$ degrees) and inherited fake \lensfit\ weights from one half onto the other. We then
compared the inherited weights to those that were originally measured. For ease of interpretation, we rescaled the \lensfit\
weights in this test to the range $w\in[0,1]$.  We found that the weight inheritance is robust, having a median residual
between the real and synthetic weights of precisely zero, being driven by the vast majority of sources residing at a true
weight of either zero or one, and being correctly assigned this limiting weight (thereby having precisely zero
residual). The scatter in the weight residuals is similarly benign, at $\sigma[w_{\rm true}-w_{\rm fake}]=0.09$ (a
perfectly random assignment of fake weights produces a scatter of approximately $0.6$). Hence, we conclude that the
weight inheritance functions appropriately.

\subsection{Tomographic binning}\label{sec:tomography}

A central aspect of weak lensing tomography is the choice of tomographic binning. In all KiDS analyses to date,
tomographic bins were defined using a set of cuts in photometric redshift ($\zb$). For the initial KiDS cosmic shear
analyses, based on optical data only, these cuts were constructed to have four bins of fixed width $\Delta \zb=0.2$
between $0.1<\zb\le0.9$ \citep{hildebrandt/etal:2017}.\footnote{We emphasise here the importance of the inequalities
used in these definitions: as the photo-$z$ estimates are discrete with finite steps of $0.01$, whether one uses
$\zb\le0.9$ or $\zb<0.9$ has a non-negligible impact on the sample definition.} With the introduction of the VIKING NIR
data and better high-$z$ performance of the photometric redshifts, a fifth (higher redshift) tomographic bin was
introduced, which used a width of $\Delta \zb=0.3$ ($0.9<\zb\le1.2$) \citep{hildebrandt/etal:2020}. These selections resulted
in tomographic bins that (for the last \kids\ analysis, see \citealt{hildebrandt/etal:2021}) contained between $2.8$
million (bin one) and $8.1$ million (bin three) sources.

This choice of tomography can, however, be shown to be sub-optimal for cosmic shear tomography signal-to-noise and
figure-of-merit in typical applications. \cite{sipp/etal:2021} advocate equipopulated bins as a better choice (over
equidistant bins), and we opted to implement this form of tomography for \kidslegacy. Details of our chosen (six)
tomographic bins, such as number densities and ellipticity dispersions, are provided in Table~\ref{tab:bins}. It should
be noted that the $\zb$ binning was chosen a priori based on the SKiLLS simulations \citep[see
Sect.~\ref{sec:simulations} and][]{li/etal:2022}. In combination with the discreteness of $\zb$, this led to bins that
are only approximately equipopulated in the \kidslegacy\ data.

\begin{table*}
  \centering
  \caption{\label{tab:bins}Properties of the six \kidslegacy\ tomographic bins and the full source sample, using our
  fiducial redshift calibration procedure.}
  \begin{tabular}{ccrrrrrrrr}
    \hline
    \hline
    Bin & Selection & \multicolumn{1}{r}{$N$} & \multicolumn{1}{r}{$n_{\rm eff}$} &
    \multicolumn{1}{r}{$\sigma_\epsilon$} & \multicolumn{1}{r}{$N_{\rm gold}$} &
    \multicolumn{1}{r}{$n_{\rm eff,gold}$} & \multicolumn{1}{r}{$\sigma_{\epsilon,{\rm gold}}$}
    & \multicolumn{1}{r}{$m_{\rm gold}$} & \multicolumn{1}{r}{$n_{\rm eff,gold}/n_{\rm eff}$} \Tstrut\\
\hline\Tstrut
    1    & $0.10<\zb\le0.42$ &   7\,442\,842 & 1.84 & 0.27 & \,7\,416\,371 & 1.77 & 0.28 & -0.0229 & 0.963 \\
    2    & $0.42<\zb\le0.58$ &   7\,382\,526 & 1.68 & 0.27 & \,7\,359\,911 & 1.65 & 0.27 & -0.0160 & 0.984 \\
    3    & $0.58<\zb\le0.71$ &   6\,803\,160 & 1.52 & 0.29 & \,6\,799\,681 & 1.50 & 0.29 & -0.0113 & 0.987 \\
    4    & $0.71<\zb\le0.90$ &   6\,880\,618 & 1.47 & 0.27 & \,6\,880\,432 & 1.46 & 0.26 &  0.0199 & 0.994 \\
    5    & $0.90<\zb\le1.14$ &   6\,477\,540 & 1.35 & 0.29 & \,6\,477\,538 & 1.35 & 0.28 &  0.0295 & 0.998 \\
    6    & $1.14<\zb\le2.00$ &   5\,963\,921 & 1.09 & 0.31 & \,5\,960\,461 & 1.07 & 0.30 &  0.0445 & 0.977 \\
    1--6 & $0.10<\zb\le2.00$ &  40\,950\,607 & 8.81 & 0.28 &  40\,894\,394 & 8.79 & 0.28 &  0.0037 & 0.983 \\
    \hline
  \end{tabular}
  \tablefoot{
    Values of $\sigma_{\epsilon}$ and $n_{\rm eff}$ are computed using
    Eqs.~C.9 and C.12 of \citet{joachimi/etal:2021}, respectively. The
    $\sigma_{\epsilon}$ values correspond to the ellipticity
    dispersion per component. $m_{\rm gold}$ corresponds to the
    multiplicative shear measurement bias.
    Statistics in the `gold' columns are computed for gold-selected
    sources using the gold weights described in Sect.~\ref{sec:
      goldweight} and include contributions from multiplicative shear
    biases, which are themselves given in the table. 
}
\end{table*}

\section{Simulations} \label{sec:simulations}

Since the first cosmic shear analysis of \kids\ \citep{hildebrandt/etal:2017}, complex image simulations have been
leveraged to calibrate shape-measurement biases \citep{fenech-conti/etal:2017}. Subsequent analyses from
\citet{wright/etal:2020b} also utilised complex simulations to estimate informative priors on redshift distribution
biases; however, these simulations were performed without an image layer. In \kidslegacy, we utilised the `\skills' suite
of image simulations \citep{li/etal:2022} to, for the first time in \kids, perform joint calibration of redshift
distribution and shape-measurement bias parameters. Additionally, \kidslegacy\ also utilises an updated version of the
MICE2 simulation \citep{fosalba/etal:2015,fosalba/etal:2015b,Crocce/etal:2015,Carretero/etal:2015} presented in
\cite{vandenbusch/etal:2020} and utilised in \cite{wright/etal:2020a}. \skills\ is a multiband image simulation based on
the SURFS dark-matter simulation \citep{Elahi/etal:2018} and Shark semi-analytic model \citep{lagos/etal:2018}, and is
constructed to match the observed multiband properties of \kidslegacy. \skills\ is specifically designed to incorporate
multi-wavelength imaging, realistic clustering, correlations between galaxies properties and environment, and
redshift-dependent shear, thereby enabling the analysis and correction of higher-order effects in shape and redshift
calibration. Conversely MICE2 is a simulated galaxy catalogue (i.e. without an image layer) derived from the MICE-Grand
Challenge simulation, which we post-processed with an analytic photometry model. The two simulations were constructed to
replicate the photometric properties of \kids\ and \viking\ data in each of the \utok\ bands as well as \lensfit\ shape
weights (beside other aspects such as shear, clustering, etc.).  

There are two additional differences between \skills\ and MICE2 that are worthy of comment. First, MICE2 covers an
on-sky area of about 5000 \sqdeg, whereas \skills\ is limited to 108 \sqdeg. Secondly, \skills\ has a much larger
redshift baseline ($0.001<z<2.5$) than MICE2, which is limited to $0.07 \lesssim z \lesssim 1.4$ and therefore did not
allow us to simulate the \kids\ data in the sixth tomographic bin (or possible high-$z$ tails of the other bins) with
high fidelity. Additionally, as mentioned above, \skills\ is used for joint shape-measurement calibration and redshift
distribution bias estimation, which enables the correction of subtle unrecognised biases that correlate errors in
shape-measurement and redshift distribution estimation (such as shear-based detection biases).  
Due to these differences, we relied on MICE2 to simulate our clustering redshift analysis (requiring the
additional area), whereas \skills\ was our primary simulation for the SOM calibration (covering the sixth bin).
Nevertheless, the two simulations are useful where they overlap, since they give us additional redundancy and allowed us to
test how our redshift estimates depend on the assumptions underlying both simulations.

\subsection{\skills}\label{sec:SKILLS}

The multiband image-simulation based \skills\ utilises imaging properties (limiting magnitudes, PSFs, etc.) sampled
directly from the \konek\ dataset \citep{kuijken/etal:2019}, such that the observational parameters are representative
of the parameter distributions therein. However, the base simulations tend to overproduce sources (relative to the data)
at low-resolution and high signal-to-noise, leading to possible systematic biases in the recovery of shape calibration
values, and which could also cause bias in redshift distribution estimates (as resolution and S/N are correlated
with colour and redshift). Therefore, in order to optimise the similarity between simulations and data,
\cite{li/etal:2022} performed an a posteriori re-weighting of the simulated wide-field sources by comparing their
abundance in a 2D space of shape-measurement S/N and source-resolution space to the KiDS wide-field data. In
\kidslegacy\ we followed the methods of \citet{li/etal:2022}, and implemented a similar re-weighting scheme to construct
our simulated calibration samples (see Sect.~\ref{sec:sim_kidz}) and corresponding wide-field samples (see
Sect.~\ref{sec:sim_wide}).

\subsection{MICE2}\label{sec:MICE2}

For \kidslegacy\ we used an updated version of the \kids-like MICE2 mocks that resembles \drfive\ and
implements an improved analytic photometry model \citep{linke/etal:2025}. We derived all the necessary calibration datasets from
the underlying base simulation. Previous \kids\ analyses have put considerable effort into analytically mimicking their observed properties as
closely as possible \citep{vandenbusch/etal:2020}, by reconstructing the samples' selection functions (typically in colour, redshift,
and/or derived properties such as stellar mass) in the simulation space. The documented spectroscopic success rate
(as a function of redshift and magnitude) was similarly included where available. \citet{vandenbusch/etal:2020} provide extensive
demonstrations of the performance of this sample construction using the
MICE2 simulation, which were used for the calibration of \konek. Generally speaking, it was difficult to faithfully reproduce the \specz\ samples
in the simulation space without ad hoc modifications to the original
selection window, and as such they were defined with a modified selection window that reproduced the expected colour,
redshift, and number density distributions seen in the data. 

Similar to \cite{vandenbusch/etal:2020}, we constructed the wide-field calibration datasets such that they match observed spectroscopic data in
sky coverage and relative overlap (e.g. between BOSS and GAMA), but additionally we decided to apply a stellar mask, which we
constructed from the real \kidslegacy\ masks by tiling the MICE2 footprint. Since we used DESI and VIPERS data for the first time in
a \kids\ clustering redshift analysis, we implemented their respective selection functions for MICE2 similar to the existing ones
for the GAMA, BOSS, and 2dFLenS samples of \cite{vandenbusch/etal:2020}. For details refer to Appendix~\ref{app:mocks}.

\subsection{Simulating the \kidz\ compilation}\label{sec:sim_kidz}

The \kidz\ spectroscopic compilation is quite different from the wide area samples described above, since it covers only
$\sim 20$~\sqdeg\ on sky and extends to both significantly higher redshifts and fainter magnitudes. In previous work
\citep{wright/etal:2020a,hildebrandt/etal:2021} we elected to apply the existing deep field selection functions in MICE2
to many distinct lines of sight (appropriately sized for the spatial extent of the data calibration samples), to
generate many realisations of the \specz\ calibration samples in the simulation volume.  This process produced $N$
realisations of the spectroscopic compilation, each of which contained different realisations of underlying sample
variance and photometric noise.  Provided enough spatial realisations, the simulations were then assumed to
span the range of possible calibration samples that could have been observed in the real Universe.  Therefore, by
calibrating our simulated wide-field sample with these realisations of the full calibration sample, we were able to
estimate an average bias (and uncertainty) that captured the range of biases that would be seen under repeated
observations of our calibration sample in different parts of the sky. 

However, this is not directly the question of relevance for our cosmic shear analysis. Rather, the calibrating sample
exhibits some particular joint distribution in redshift and colour, and we wish to identify the bias that is introduced
to our analysis due to that specific joint distribution.  In previous \kids\ work using MICE2 (whose light
cone covers a full octant of simulated sky), \cite{wright/etal:2020a,hildebrandt/etal:2021} used $N=100$ lines of sight
to estimate the uncertainty on the redshift calibration procedure. Should our observed calibration sample be an outlier
in the distribution of all possible sample variance and photometric noise realisations, then there is only a small
chance that such a realisation exists in a sample of $100$ lines of sight.  This is not formally a problem, but does
decrease the interpretability of our cosmological posteriors somewhat. 

Hence, in \kidslegacy\ we shifted the philosophy of our simulated analyses to focus on the issue of discerning the
bias from the calibration sample that we actually have, rather than marginalising over the uncertainty from all possible
calibration samples. This required a change in implementation of the construction of the calibration samples in the
simulations. The new method of constructing realistic mock calibration samples (see Sect.~\ref{sec:sim_match}) was applied to
both \skills\ and MICE.

\subsubsection{Sample matching}\label{sec:sim_match}
As described in Sect.~\ref{sec:sim_kidz}, the procedure for generating redshift calibration samples in simulations for
\kidslegacy\ was updated to produce more accurate estimates of the redshift calibration bias present in the actual
distributions of calibrating spectra available to us. This involved directly replicating the distribution of available
calibrating spectra in multi-dimensional colour, magnitude, redshift, and \photoz\ space.

We performed the multi-dimensional matching using the
\texttt{galselect}\footnote{\url{https://github.com/jlvdb/galselect.git}} python module. The module takes two
catalogues: a `candidate' catalogue of potential sources, and a `target' catalogue, which we wanted to reproduce. The module
also takes a list of input features (such as colours and/or magnitudes), and a true-redshift designation for the two
catalogues. In practice, we performed the matching in \kidslegacy\ using our ten-band magnitudes as the input features. 
With this information, the algorithm performs a brute-force search around each entry of the target
catalogue to choose the best-matching candidate catalogue object. This brute-force search first involves truncating the
candidate catalogue in a thin slice of true redshift around the target source redshift. This in effect forces the
resulting matched catalogue to have the exact \nz\ of the target catalogue, agnostic to the quality of the matched
features. The feature match is then performed by computing the Euclidean distance (in the
$N$-dimensional feature space) between all candidate objects and the target source. The best matching object is then
chosen to be the candidate with the lowest Euclidean distance or (optionally) the candidate with the lowest Euclidean
distance that has not previously been matched to a target source (i.e. allowing or not allowing candidate objects to be
duplicated, respectively). 

The algorithm therefore contains two primary options that are arbitrarily chosen by the user. Firstly the size of the
window in true-redshift surrounding each target source that is used to define the possible candidate objects; and
secondly features that are used to define the matching. In Sect.~\ref{SOM: matching} we outline the influence of
these options on the constructed calibration samples. 

It should be noted that this algorithm, while yielding close-to perfectly matched redshift, colour, and magnitude
distributions, does not necessarily also yield a sample with realistic clustering properties. Indeed, we believe that
some of the samples constructed this way might have pathological clustering properties. Therefore, we decided to not use
the matching algorithm for creating the wide-field samples used in the clustering redshift analysis on MICE2 and revert
to the more traditional method of directly replicating the spectroscopic target selections there (see
Sect.~\ref{sec:MICE2}).

\subsubsection{Matching to wide-field sources}\label{sec:sim_wide}

One problem with the implementation of our matching approach for construction of the calibration samples in our
simulations is that, if there are any systematic differences in the colour-redshift space between the simulations and
the data, the matching algorithm will introduce a systematic discrepancy between the colour-redshift relation in
the calibration- and wide-fields. 
To mitigate this possible effect, we implemented a similar matching algorithm between the data and simulation wide-field
samples. However, this implementation cannot, of course, use true redshift as a basis (as in Sect.~\ref{sec:sim_match}).
Instead, we aimed to reproduce the wide-field sample in the simulations by matching sources, again by colour and
magnitude, in discrete bins of \photoz, ensuring a perfect match of the \photoz\ distributions. 

The algorithm proceeds simply by selecting all sources from both the wide-field samples on the data and simulations that
reside at a particular (discrete) value of \photoz. These samples are then matched to one another using a
$k$-nearest-neighbour method, and all simulation sources are tagged with the number of data-side sources that were most
closely matched to them. This allowed us to construct frequency weights for all sources in the simulated wide-field
sample, to more accurately reflect the observed distribution of sources in colour and \photoz.  The resulting
frequency-weighted wide-field sample is then used for calculation of redshift distributions and bias parameters.  

\section{Direct calibration with SOMs}\label{sec:SOM} 

For all cosmological analyses with \kids\ since \cite{hildebrandt/etal:2017}, the fiducial estimation of redshift
distributions and their calibration has been performed via some implementation of direct calibration
\citep{lima/etal:2008}. \cite{wright/etal:2020a} presented an implementation of direct calibration using SOMs that has
been utilised in all cosmological analyses with \kids\ since 2020. In \kidslegacy, we also implemented a version of
direct calibration with SOMs as our fiducial redshift estimation method, however, with a number of modifications not
present in previous studies. 

The calculation of redshift distributions for \kidslegacy\ was performed within the
\cosmopipe\footnote{\url{https://github.com/AngusWright/CosmoPipe}}  pipeline, described primarily in
{\citet{wright/etal:2025b}} and used in an earlier form by \citet{wright/etal:2020b} and \citet{vandenbusch/etal:2022}.
Within \cosmopipe, redshift distribution estimation was achieved using a sequence of processing functions. Crucial
differences in the redshift distribution estimation procedure, compared to that implemented in previous analyses of
\kids, are: the use of one SOM per tomographic bin (Sect.~\ref{sec: tomosom}), the use of gold weight rather than gold
class (Sect.~\ref{sec: goldweight}), and additional weighting on the calibration sample to account for prior-volume
effects (Sect.~\ref{sec:priorweight}). 

\subsection{Tomographic SOM construction}\label{sec: tomosom} 

In their SOM implementation of direct calibration for \kids, \cite{wright/etal:2020a} trained a $101\times101$ cell SOM
on the full KiDS+VIKING-450 \citep{wright/etal:2019} calibration sample of $25\,373$ sources, corresponding to roughly
two calibrating sources per-cell on average. This SOM was then utilised to compute individual tomographic bin redshift
distributions by subsetting the calibration sample (using the photometric redshift limits that define the tomographic
bins) prior to the computation of direct calibration weights (DIR; see the beginning of their Section 4). Motivations
for this choice are documented in \cite{wright/etal:2020a}, and focus (in particular) on systematic biases that occur
when constructing \nz\ using the full calibration sample rather than tomographically binned calibration samples. This
process, however, resulted in a significant decrease in the number of sources that were calibrated by spectra in the
wide-field sample (as much as a 30~\% reduction in the available number of sources in some tomographic bins). To
circumvent this issue, the SOM cells were then merged using full-linkage hierarchical clustering to maximise coverage of
the wide-field sample while maintaining a robust estimate of the redshift distribution. These merged groups of cells
were then used in the computation of the DIR weights. 

One caveat of the above procedure is that the number of cells assigned to regions of the colour-magnitude space
dominated by the individual tomographic bins is non-uniform: tomographic bins with relatively fewer calibrating spectra 
receive fewer cells and less coverage in the combined SOM. This was not a problem for previous studies that used \kids;
however, in this study we introduced a new, higher redshift tomographic bin. This tomographic bin is both noisier (in
terms of photometric properties) and has relatively fewer calibrating spectra than its lower redshift counterparts
despite the increased number of spectroscopic calibration sources: $126\,085$, more than the previous \kids\ calibration
sample by a factor of roughly 5. 

Therefore, in order to make optimal use of this larger calibration sample and accurately calibrate the higher-redshift
tomographic bin, we opted for training one SOM per tomographic bin.  This ensures that each tomographic bin contains the
same number of cells in the training, and avoids the limitations that can be imposed by utilising a single SOM for
calibration of the entire shear sample. The settings for the SOM training are summarised in Table~\ref{tab:SOMparams}.
In particular, we note that the change to tomographic SOMs is accompanied by a reduction in the SOM size, from
$101\times101$ to $51\times51$, to ensure similar cell population statistics when between tomographic and
non-tomographic SOMs.  

\begin{table}
    \centering
    \caption{Fiducial parameters for SOM construction in
      \kidslegacy.}
    \begin{tabular}{cc}
    \hline\hline
    Parameter & Value \Tstrut\Bstrut\\ 
    \hline 
    Training sample & Tomographic calib. sample \Tstrut\\
    SOM realisations & 10 \\
    Training expression & All colours \& $r$-band total \\ 
    Dimension & $51\times 51$\\
    Topology & toroidal  \\
    Cell type & hexagonal  \\
    Data magnitude limits & $r\in \left[20, 24.5\right]$ \\
    Calibration weighting & Shape \& prior volume \\
    Training iterations & 100 \Bstrut\\  
    \hline
    \end{tabular}
    \label{tab:SOMparams}
    \tablefoot{The term `All colours' refers to all non-redundant combinations of magnitudes that we are able to
    construct from the ten-band photometry, including the magnitude difference computed between the two $i$-band passes.
    As this difference does not encode SED shape information, nevertheless, we also tested the results excluding the
    $i$-band difference, finding them unchanged with respect to the fiducial case.}
\end{table}

As a quantitative demonstration, we computed the tomographic-bin coverage statistics for a single $101\times101$ SOM
trained in the same manner as for previous \kids\ analyses. In such a SOM, the partitioning between individual
tomographic bins is relatively good, with all tomographic bins covering $11-21$~\% of cells (a perfect equipartition
would correspond to approximately 15~\% after accounting for sources beyond the tomographic limits included in the
training). Nonetheless, there is a factor of $\sim2$ difference in coverage between some bins, with bins four and five
dominating ($19.5\%$ and $21.2\%$ of cells, respectively). This is expected, as the unweighted \nz\ of the calibration
sample peaks in the region $0.6<z<1.1$, where the bulk of bins four and five reside. In order to avoid this
over-representation of the SOM manifold by bins four and five and give equal weight to all bins, we moved to individual
SOMs for each bin.

\subsection{Gold class versus gold weight}\label{sec: goldweight} 

In the SOM redshift calibration implemented by \cite{wright/etal:2020b}, the authors introduced the `gold' selection to
the cosmic shear analyses. This selection flagged and removed sources that resided in parts of the colour-magnitude
space that did not contain calibrating spectra. This gold-class selection improved the robustness of recovered
cosmological constraints, by removing sensitivity of the recovered cosmology to systematic misrepresentation of
calibrating spectra, which resulted in potential redshift biases, and which are a natural outcome of the wildly different
selection functions between samples of galaxies from spectroscopic and wide-field imaging surveys. 

In the establishment of the gold class, \cite{wright/etal:2019} demonstrated that repeated construction of the gold
class led to changes in the effective number density of sources (per tomographic bin) at the level of $\leq 3$~\%,
indicating that the gold selection was robust under repeated analysis. However, repeated end-to-end analyses of
KiDS-1000 (within the new \cosmopipe\ pipeline) showed more random noise in the recovered cosmological constraints than
would naively be expected from a 3~\% change in the sample (with all other analysis aspects being equivalent to their
KiDS-1000 counterparts).  Investigation of this effect demonstrated an unrecognised feature of the gold selection.
While the effective number density of sources per bin is stable under repeat analysis, the assignment of a gold flag to
the sources themselves varies to a much higher degree. For example, repeated computation of the gold class will
consistently classify 15~\% of sources as non-gold in a given tomographic bin, but precisely which 15~\% of sources are
removed may vary from realisation to realisation. This has the effect of increasing the independence of the samples
that are used in the end-to-end reruns of the cosmic shear data vector, and therefore increases the noise in estimated
cosmological parameters between reruns.  Testing on KiDS-1000, for example, demonstrated that the variation between
shape-noise realisations implicit to the changing gold class could lead to variations in marginal constraints of $S_8$
at the level of $\lesssim 0.5\sigma$; much larger than one would expect from an apparent $\sim 3\%$ change in the
sample. 

The cause of this effect is primarily photometric noise and the random nature of the SOM training, which combine to
produce highly stochastic assignment of spectra to cells under any one training. While such run-to-run variation is not
necessarily a problem a priori (the issue described above is rather with our assumption that the samples are consistent
between trainings), we nonetheless sought an analysis alternative that reduced the sensitivity of our cosmic shear
measurements to the training of an individual SOM. The simplest alternative is to perform the SOM training many times,
and utilise the distribution of gold-class assignments as a weight in the final cosmological analysis: the
`gold weight'. 

\begin{figure*}
  \centering
  \includegraphics[width=0.29\textwidth]{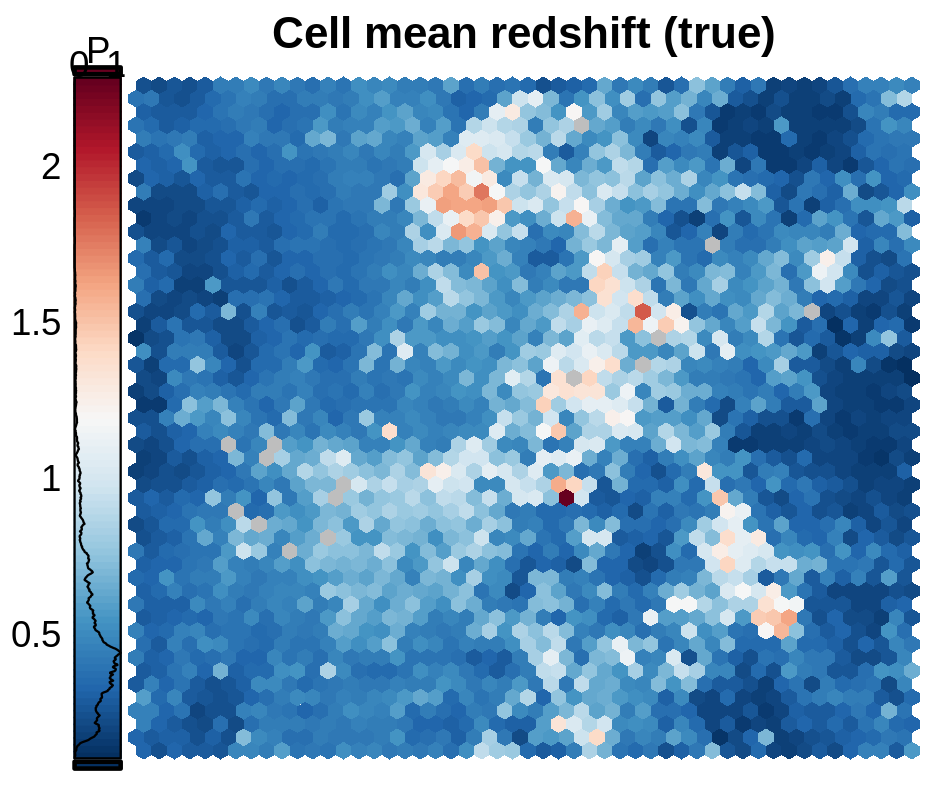}
  \includegraphics[width=0.3\textwidth]{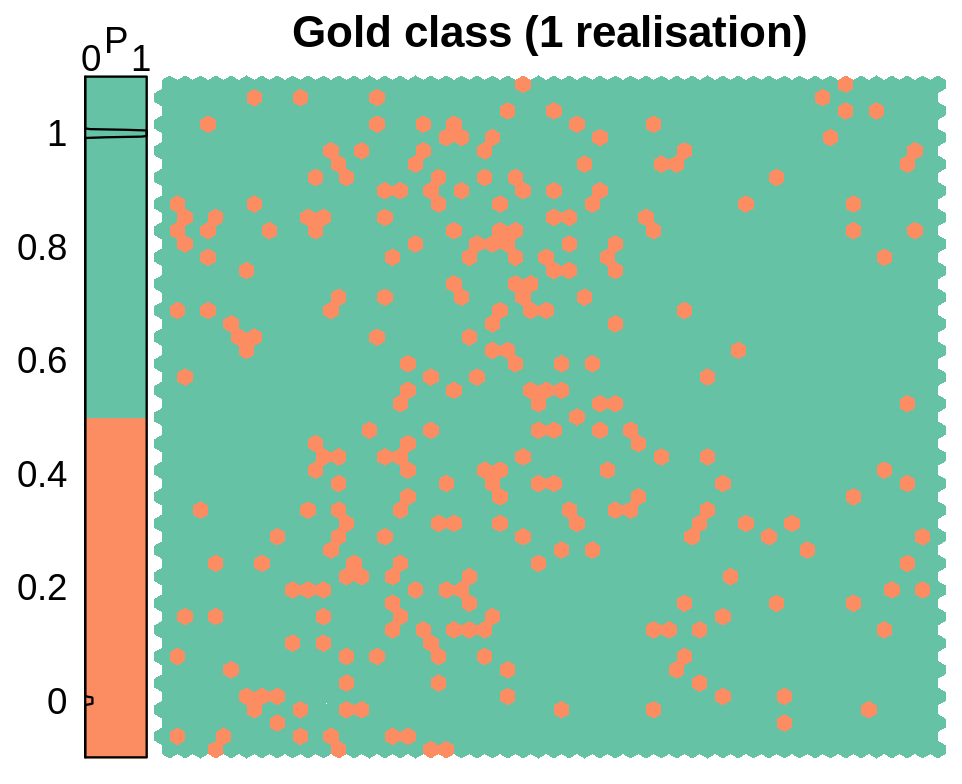}
  \includegraphics[width=0.3\textwidth]{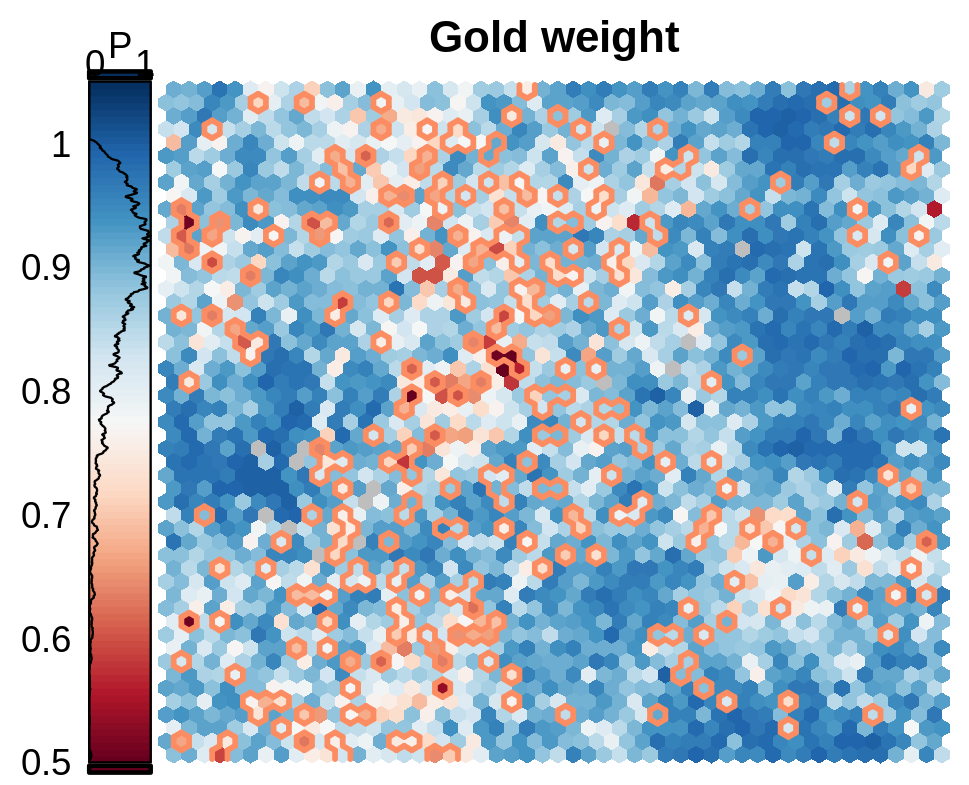}
  \caption{Comparison between gold-class and gold-weight definitions. Here, a SOM trained on tomographic bin one in our
  \skills\ simulation is coloured by the true mean redshift of each cell ({\em left}), the gold-class definition of each
  cell under a single realisation ({\em centre}), and the gold weight of each cell after ten realisations. In the
  gold-weight panel, the cells, which are assigned a (highly stochastic) zero gold class in our single realisation, are highlighted with an
  orange border. These cells are assigned a wide range of final
  gold-weights, highlighting the stochasticity of the gold-class
  definition and the superiority of the gold-weights. In each colour bar, the PDF of cell values is shown.}\label{fig:class_vs_weight}
\end{figure*}

We computed the gold weight by training $N_{\rm repl}$ SOMs (either using the full sample or individual tomographic bin
samples; see Sect. \ref{sec: tomosom}) and calculated the gold class of all sources per bin for each of these SOMs. The
gold weight was then defined as
    \begin{equation} 
      W^{\rm gold}_i = { \sum_{j \in N_{\rm repl}} g_{i,j} \over N_{\rm repl} }\;,
    \end{equation}    
where the $g_{i,j}$ are the $0/1$ gold classifications assigned to each source $i$ under realisation $j$ of the SOM
training. Using this gold weight, we were able to construct \nz\ that are less sensitive to the randomness of any single
gold-class assignment, and to construct data-vectors that are more consistent under end-to-end reruns of the analysis
pipeline.  This has the primary benefit of creating less statistical noise in repeated analyses of KiDS, leading to a
more robust legacy data product. An additional benefit of gold-weighting is that it eliminates the requirement for the
hierarchical clustering of SOM cells, which was performed by \citet{wright/etal:2019} to increase the fraction of
positive gold-class assignments for wide-field sources.  Figure~\ref{fig:class_vs_weight} demonstrates the benefit of
gold weight over gold class visually. The gold-class definition is highly stochastic, as cells that are classed as gold
in a single realisation are assigned a wide range of gold weights after many realisations. 

The distribution of gold weights computed for our fiducial simulations in \kidslegacy\ are shown in Figure \ref{fig:
goldweight}. It is apparent from the figure that the gold weight per source varies strongly per tomographic bin; however,
the behaviour is qualitatively similar in the most relevant aspects: all bins have a peak in their gold-weight
probability distribution function (PDF) at
unity (implying that sources are typically consistently classed as gold under realisations of the SOM), and very few
sources have a gold weight of zero (suggesting that it is rare for sources to be consistently unrepresented in the
calibration compilation under realisations of the SOM). This result is consistent with the conclusion that the
variability in gold assignment is driven by photometric noise. 

\begin{figure}
  \centering
  \includegraphics[width=\columnwidth,trim={0mm 10mm 0mm 16mm},clip]{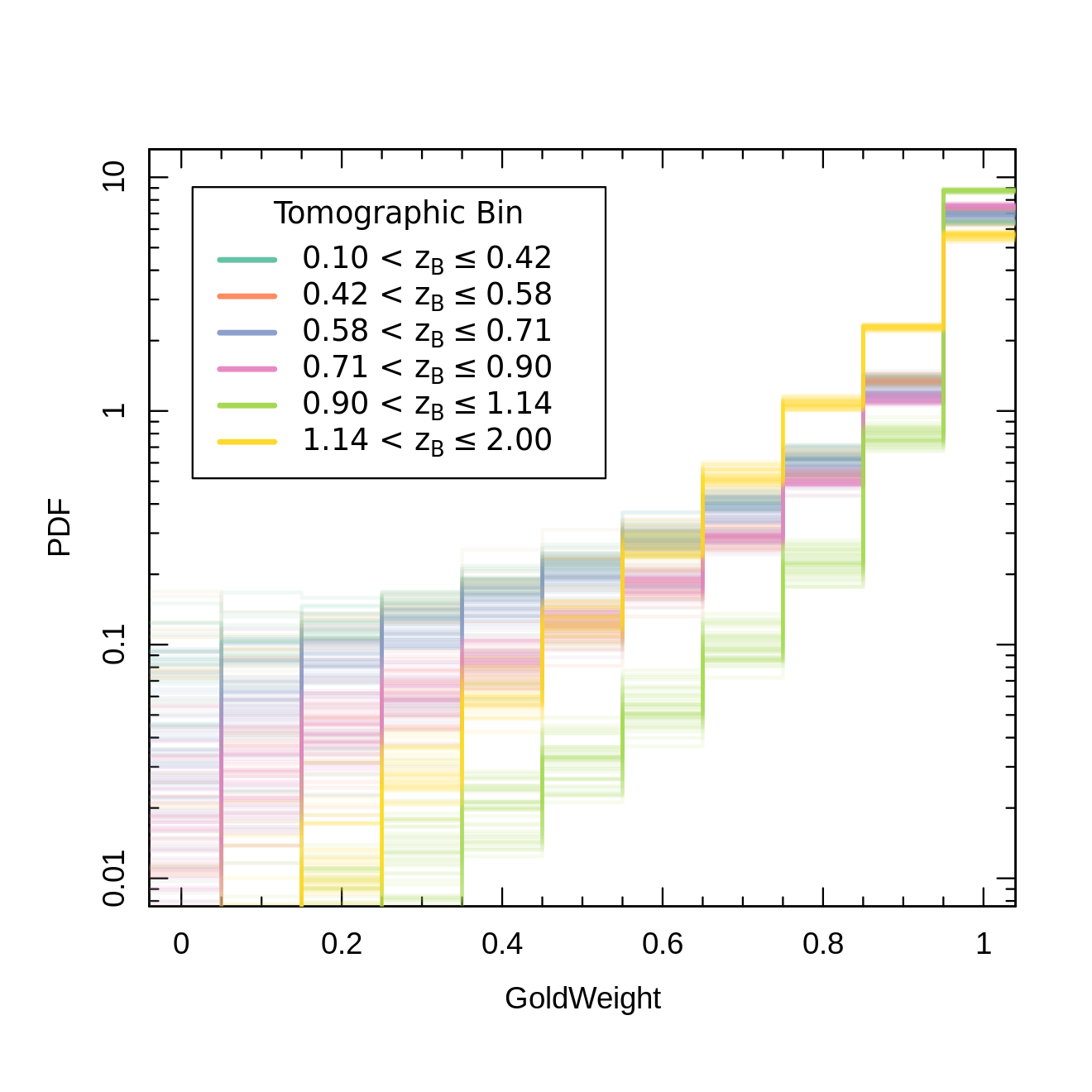}
  \caption{Distributions of gold weight per tomographic bin for our fiducial \skills\ simulations. Individual lines show the scatter in the
  gold-weight PDFs under different realisations of our spectroscopic calibration samples. The tomographic bins show
  qualitatively similar behaviour: many sources are consistently classed as gold under all realisations of the SOM 
  ($w_{\rm gold}=1$), and very few sources are consistently classed as not-gold under all realisations ($w_{\rm
  gold}=0$).  }\label{fig: goldweight}
\end{figure}

\subsection{Prior redshift weight}\label{sec:priorweight} 

The SOM implementation of the direct calibration method is designed to perform two primary tasks: re-weight the colour
space of the calibration sample to better represent the wide-field sample, and flag wide-field sources for removal where
this correction is not possible (i.e. the gold-weighting).  These corrections assume, however, that the probability
distribution of redshift at a given colour in the calibration and wide-field samples are identical. Such an assumption
is easily violated in the process of spectroscopic redshift acquisition, where two galaxies with different redshift but
the same broadband colours (i.e. those with colour-redshift degeneracy) have different spectroscopic redshift success
rates (e.g. one galaxy shows the [OII] doublet in the optical and the other does not). Such a selection in the
successful acquisition of spectroscopic redshifts has been shown to lead to pathological biases in vanilla direct
calibration implementations \citep{hartley/etal:2020}.  

Even more simply, however, this assumption is also easily violated when the samples are constructed from vastly
different selection functions \citep[e.g.][]{gruenbrimioulle:2017}.  For example, two simple magnitude-limited samples
constructed from different magnitude limits will probe different redshift baselines. If the deeper sample has access to
galaxies that are colour-degenerate with galaxies in the shallower window, then the redshift distribution at fixed
colour will be unimodal for the shallow sample and multi-modal for the deeper sample. 

Correcting for this effect is complicated, as it requires one to know the distribution of redshift for the target sample
of galaxies (which is our desired end-product of the SOM calibration process).  In \kidslegacy, we performed a first
order correction using an a priori estimate of the wide-field sample redshift distributions (see below) and removed
significant differences between this wide-field estimate and the (known) redshift distribution of the calibration
sample. 

\begin{figure*}
    \centering
    \includegraphics[width=\linewidth]{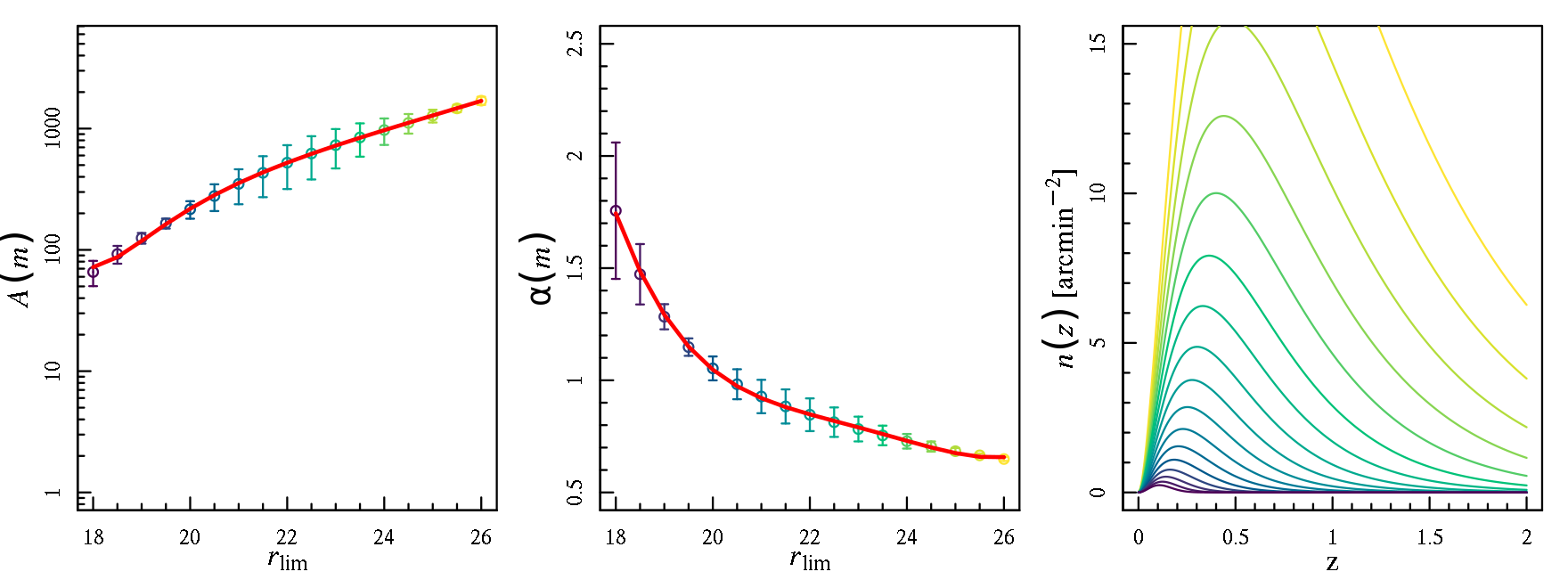}
    \caption{Model parameters and the resulting analytic \nz\ for
      samples defined as magnitude-limited (in the
      $r$ band) from our noiseless SURFS+Shark light cone. 
      The left and centre panels show the free parameters from Eq.~\eqref{eq:analyticnz}, 
      as a function of the $r$-band magnitude limit, including polynomial fits. The right panel shows the analytically 
      estimated \nz\ for each of the models parameters in the other two panels.}
    \label{fig:analytic_nz}
\end{figure*}

To perform this correction, we required an estimate of the true redshift distribution of the cosmic shear wide-field
galaxy sample. To this end we began by constructing an analytic expression for the redshift distribution of an arbitrary
magnitude-limited sample. Using the raw $108$~\sqdeg\ SURFS-Shark light cone (see Sect.~\ref{sec:simulations}), which
contains noiseless SDSS and VIKING fluxes, we constructed samples of galaxies cut to various magnitude limits in a range
of photometric bands. We then fitted each of the resulting galaxy samples with the function: 
\begin{equation}\label{eq:analyticnz}
  N(z,m)=A(m) \, z^2 \exp{\left(-(z/0.1)^{\alpha(m)}\right)}\;,
\end{equation}
where $A$ and $\alpha$ are free parameters. We then modelled $A(m)$ and $\alpha(m)$ with a fourth-order polynomial.
This allowed us to construct an analytic redshift distribution for a sample of galaxies that is magnitude-limited (in
true flux) between the 18$^{\mathrm{th}}$ and 27$^{\mathrm{th}}$ magnitude in any band from $u$ to $Z$.  An example
showing the estimated model parameters, the polynomial fits, and the resulting analytic \nz\ is given in
Fig.~\ref{fig:analytic_nz}. 

We subsequently constructed prior volume corrective weights for the \kidslegacy\ calibration sample using this analytic
prescription. To do this we first selected a magnitude-limited sample that we found closely mimics (in terms of the
redshift distribution) the expected true selection function of our wide-field shear dataset; a process that is
complicated by the various complex lensing selections (and shape weights) that are applied to the wide field lensing
sample of galaxies.

\begin{figure}
    \centering
    \includegraphics[width=\linewidth]{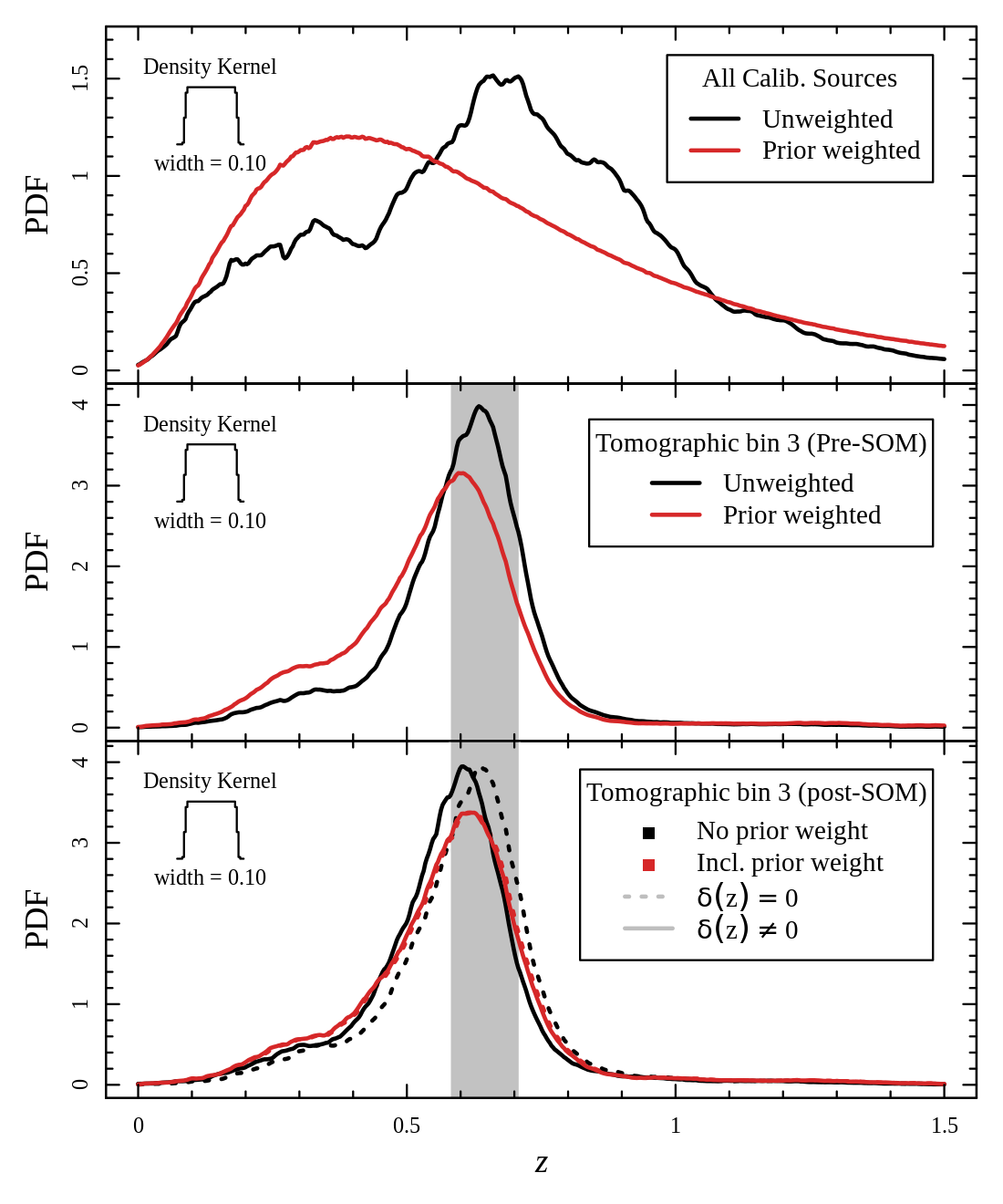}
    \caption{Impact of prior-volume weights on the (pre-SOM) \nz\ of the full spectroscopic compilation (upper panel),
    and on those galaxies in the compilation that end up in tomographic bin three both prior to the SOM re-weighting
    (middle panel) and after the SOM re-weighting (bottom panel).  The systematic effect that the prior weight imparts
    on the tomographic bin is particularly clear prior to re-weighting by the SOM to match the wide-field sample but is
    somewhat diluted by the SOM re-weighting process. Nonetheless, there is a residual difference after the SOM
    re-weighting in the bottom panel.  This difference is exacerbated if one does not have high quality image
    simulations to calibrate the underlying \nz\ bias (dashed lines in the bottom figure; discussed further in \protect
    Sect.~\ref{results: simulations}), as the non-prior weighted \nz\ estimate contains a significant systematic bias
    ($\langle \delta z\rangle \approx 0.03$), whereas the prior weighted \nz\ estimate has no bias ($\langle \delta
    z\rangle \approx 0.004$). }
    \label{fig:specz_priorwgt_tomo}
\end{figure}

We chose to use a sample that is magnitude-limited in the $r$ band, at $20\leq r\leq23.5$. 
The bright-end magnitude limit was chosen because of the selection performed by \lensfit that rejects galaxies at $r<20$. 
The faint-end limit was chosen due to the lensing weights returned by \lensfit, which are strongly magnitude-dependent: at
$r\approx 23.5$ the lensing weight is roughly half its maximum. With our analytic estimate of the wide-field sample
redshift distribution, we then defined
the corrective prior volume weights as the ratio of the redshift distribution PDFs $P_w(z)/P_c(z)$, where $w$ and $c$
refer to the analytic wide-field sample and the data calibration sample, respectively. The impact of the prior volume
weights on the total \nz\ of the spectroscopic compilation, and also for an example tomographic bin, are shown in
Fig.~\ref{fig:specz_priorwgt_tomo}. The figures show the distributions both before and after SOM weighting, and when
including our simulation-informed estimates of \nz\ bias (see Sect.~\ref{results: simulations}). It is clear
from the figure that the prior volume weights have a systematic effect on the relative weight of individual calibrating
sources as a function of redshift and, perhaps more importantly, that this manifests as a shift in the entire
pre-weighting \nz\ for some tomographic bins. After weighting by the SOM, however, the differences are less pronounced,
indicating that some of the difference between the redshift distributions is absorbed by the colour-cell DIR weights
(thanks to the colour-redshift relation). Finally, after incorporation of our simulation \nz\ bias estimates the
differences are reduced even further (see Sects.~\ref{sec:bias} and \ref{results: simulations}), demonstrating the
importance of having realistic simulations to correct for methodological bias in \nz\ estimation and calibration. 

\subsection{Redshift distribution bias estimation}\label{sec:bias}

Calibration of the redshift estimation process (specifically the derivation of tomographic bin redshift bias parameters) 
in \kidslegacy\ was performed by implementing the redshift distribution estimation pipeline on our various simulated
datasets, which are designed to mimic the observed data as accurately as possible. With the estimated redshift
distributions and the known true (weighted) redshift distribution of the source samples, we then computed the bias of
each estimated redshift distribution as 
\begin{equation}\label{eqn:som_bias}
\delta z = \hat\mu_z - \mu_z\;,
\end{equation}
where $\mu_z$ is the (shape- and gold-) weighted mean true redshift of the wide-field sample, and $\hat\mu_z$ is the
estimated mean redshift of the wide-field sample, computed directly from our weighted calibration sample\footnote{For
our fiducial \nz\ binning, $\Delta z = 0.05$, the primary probability mass of each \nz\ is sampled by ten or more bins.
This means that the difference introduced when computing the sample mean redshift vs \nz\ expectation is negligible.}.
We typically perform this measurement using many realisations of the calibration samples, which produces many estimates
of $\hat\mu_z$ (and, because of the gold selection or weighting, possibly many different $\mu_z$). Our final
quoted biases are the arithmetic means of the biases estimated per tomographic bin and pipeline setup ($\langle
\delta z \rangle$). The population scatter of the biases in these realisations is also a relevant consideration, and is
quoted as $\sigma_{\delta z}$. We note in particular that these uncertainties are smaller than those in previous KiDS
analyses, due to the change in simulation philosophy described in Sect.~\ref{sec:sim_kidz}.
 
\section{Clustering redshift methodology} \label{sec:CC}

As a complementary approach to test and validate the SOM \nz\, we used clustering redshifts \citep[e.g.][]{newman:2008}
following previous \kids\ work
\citep{hildebrandt/etal:2017,hildebrandt/etal:2020,hildebrandt/etal:2021,morrison/etal:2017,vandenbusch/etal:2020}.  The
\drfive\ analysis presented here is an evolution of these previous studies, adding more area for the measurements of the
CCs and, more importantly, expanding the suite of external spectroscopic surveys used for the calibration.  The
dedicated \kidz\ data allowed us to measure CCs with VIPERS, providing additional constraints in the range
$0.6<z\lesssim 1.2$. Recently, the DESI Early Data Release provided a number of additional galaxies with spectroscopic
redshift estimates extending to even higher redshifts and overlapping with the \kids\ main survey area in six
rosette-shaped DESI pointings. Together these advances allowed us to validate the SOM \nz\ for the first five
tomographic bins (bin six is only partly covered) without referring to deep, pencil-beam surveys \citep[as we still had
to do in][]{hildebrandt/etal:2021}, significantly decoupling the clustering redshift from the SOM approach in terms of
calibration data. One caveat, though, is VIPERS, which is used for both CCs and SOM \nz\ calibration. However we found
that the overall contribution of VIPERS both SOM and CC \nz\ calibrations is subdominant (see Sect.~\ref{sec:cc_fid})
and so the independence of the methods is maintained.

\subsection{Correlation measurements}

\kids\ clustering redshifts are estimated with the versatile, public code
\yaw\footnote{\url{https://pypi.org/project/yet-another-wizz/}} \citep[YAW,][]{vandenbusch/etal:2020}, which is based on
concepts already introduced by \citet{schmidt/etal:2013} and \citet{morrison/etal:2017}. In particular we used the
publicly available version 2.6.0, which differs from the versions used in previous publications
\citep[e.g.][]{hildebrandt/etal:2021,naidoo/etal:23} in a few ways.

The code now generates $N$ spatial regions per calibration sample based on
$k$-means clustering of sky coordinates (e.g. using random catalogues) instead of splitting the data into
the $N$ individual square-degree pointings spanned by each dataset. These regions are used to estimate the data
covariance via a spatial jackknife (previously using bootstrap).  This empirical data covariance was tested against
analytical models for the covariance matrix based on halo occupation distributions and a halo model approach for a
different calibration dataset derived from MICE2. While there is good general agreement between the features of the
empirical and the analytical covariance, we opt to rely on the jackknife method due to the highly non-linear regime of
the clustering measurements, possible uncertainties in the halo occupation distribution, and non-Limber effects
in the connected non-Gaussian terms \citep[for details we refer to Section 9.2 of][]{reischke/etal:2024}. Regardless,
the agreement with the analytic covariance serves as a good cross-check for the empirical jackknife covariance that we
use throughout the clustering redshift analysis.

In addition to the above, the code now measures pair counts across the boundaries of these spatial regions\footnote{While
counting pairs only within the same region can have an effect on the overall correlation amplitude, it has no effect on
any of our previously published results because this overall amplitude is later normalised.}  and the Landy-Szalay
estimator \citep{landy/szalay:1993} is used for all auto-correlation measurements. Finally, in the case of the CCs,
which use the \citet{davis/etal:1983} estimator, only one random catalogue is needed. Hence, one can decide whether to
use a random catalogue for the spectroscopic or \kids\ data. We decided to use random catalogues for the spectroscopic
data instead of the \kids\ data in those cases, since most of the spectroscopic surveys provide well-established random
catalogues that take into account and correct for a lot of systematic effects.

\subsection{Fiducial analysis setup}\label{sec:cc_fid}

We measured angular correlations in a single bin of fixed transverse physical separation between $0.5 < r \leq
1.5~\mathrm{Mpc}$ in 32 linearly spaced redshift bins in the range $0.05 \leq z < 1.6$.\footnote{ These scales are
larger than what was used in previous KiDS clustering redshift analyses. See Sect.~\ref{sec:CC_true_nz} for a
justification.  } For MICE2, the upper redshift limit is reduced to $z_{\rm max} = 1.4$.

One of the main goals of any clustering-redshifts setup is to suppress and mitigate the redshift evolution of the galaxy
bias of the calibration data and the target wide-field dataset. Following the notation of \citet{vandenbusch/etal:2020},
these biases can, under certain assumptions, be expressed in terms of the amplitudes of the angular autocorrelation
functions of our spectroscopic reference sample, $w_{\rm ss}(z)$, and the wide-field photometric sample, $w_{\rm
pp}(z)$.  Then, the true (unknown) redshift distribution can be written as
\begin{equation}
  n_{\rm p}(z)
    = \frac{w_{\rm sp}(z)}{\sqrt{\Delta z^2 \, w_{\rm ss}(z) \, w_{\rm pp}(z)}}
    = \frac{\ccnz}{\sqrt{w_{\rm pp}(z)}} \; ,
  \label{eq:cc_nz}
\end{equation}
where $w_{\rm sp}(z)$ is the CC amplitude between the spectroscopic and photometric samples, $\Delta z$ is the bin width
of the CC measurements, and \ccnz\ denotes our estimated \nz\ from the CC method.

In our fiducial analysis we chose to only correct for the calibration data bias; that is, we effectively measure the
numerator of the right-hand side of Eq.~(\ref{eq:cc_nz}). This is achieved by measuring the angular auto-correlation
function of the calibration sample in the same redshift and distance bins as the CC function. This approach has been
shown to successfully remove the redshift evolution of the galaxy bias of that sample. However, a similar treatment is
not possible for the \kidslegacy\ source samples. The impact of the unknown galaxy bias of that sample data is partly
mitigated by the fact that we bin the data tomographically \citep{schmidt/etal:2013}, and it has been shown previously
to be sufficiently small that we can neglect it in our following analysis \citep[see
e.g.][]{hildebrandt/etal:2021,vandenbusch/etal:2020}.

\begin{figure}
  \centering
  \includegraphics[width=\columnwidth]{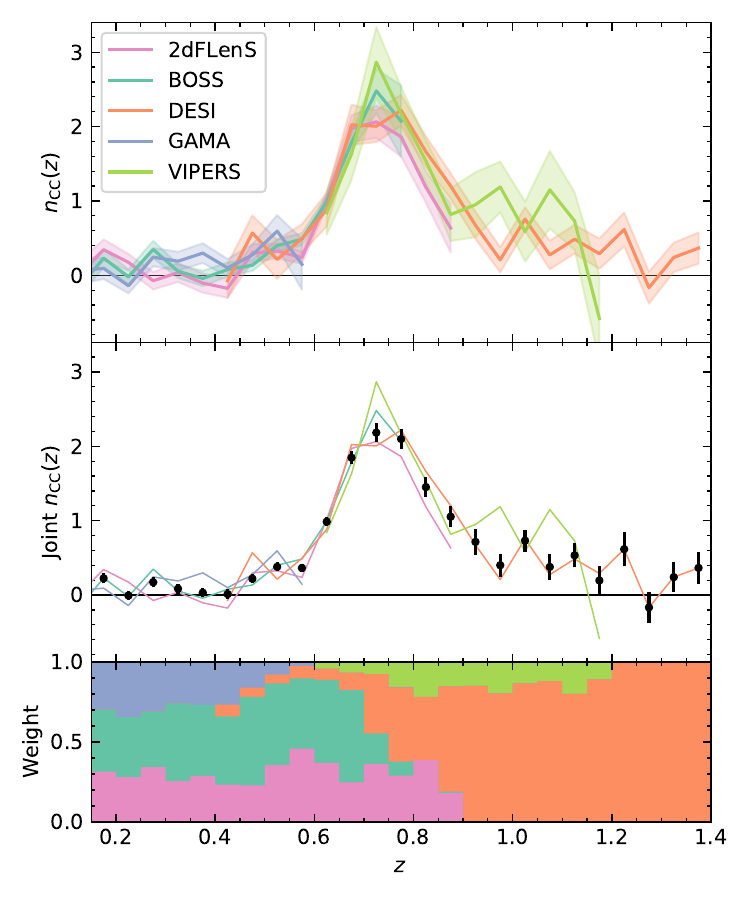}
  \caption{
    Example of the inverse-variance weighted combination of CCs computed from the fourth tomographic bin of MICE2. The
    top panel shows the individual measurements from each calibration sample, the middle panel the weighted average, and
    the bottom panel the relative weight of each sample as a function of redshift.
  }
  \label{fig:inv_var_merge}
\end{figure}

Finally, we produced a joint redshift estimate from all five calibration samples for which we measured the correlation
functions independently. We obtained this final estimate by computing the inverse-variance weighted average of the
bias-corrected correlation measurements of all reference samples. This produces an optimal redshift estimate that
reflects the redshift-dependent relative contribution of each correlation measurement to the overall redshift estimate
and accounts for the different statistical power each calibration sample has at any given redshift
(Fig.~\ref{fig:inv_var_merge}). In general, the joint CCs were found to be dominated by BOSS, GAMA, and 2dFLenS at $z
\lesssim 0.7$ and by DESI at high redshifts, whereas VIPERS had an overall small contribution due to its limited
area and number density.

\subsection{Adaptation of signal-to-noise in MICE2}\label{sec:noise_adapt}

\begin{figure}
  \centering
  \includegraphics[width=\columnwidth]{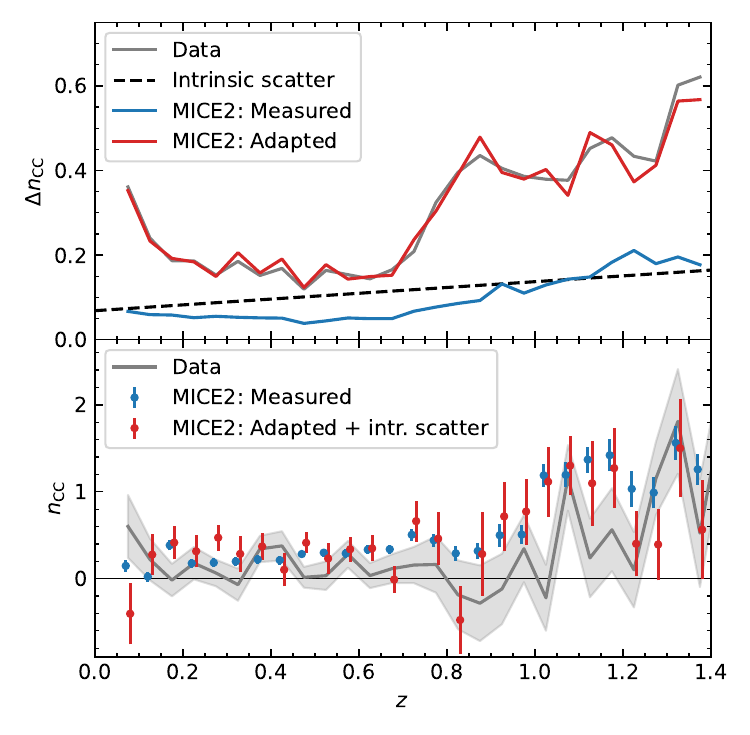}
  \caption{
    One realisation of the CC measurements from MICE2 in the sixth
    tomographic bin after noise adaptation (Sect.~\ref{sec:noise_adapt}) and the addition of intrinsic
    scatter (Sect~\ref{sec:shift_fit}).
    The top panel compares only the uncertainties of \ccnz\ from the data (grey) and MICE2 before (blue) and after adaptation
    (red). The dashed black line indicates the fitted intrinsic
    scatter from the measurements on the data ($f=0.14 \pm 0.04$; see Eq.~\ref{eq:errterm}).
    The bottom panel shows the measured \ccnz\ from MICE2 (blue points) compared to the data (grey line). The red data points
    represent the MICE2 measurements after adapting the noise (up-scaling errors and perturbing values) and adding the intrinsic
    scatter (only perturbing valu-es) obtained from the data.
  }
  \label{fig:adapt_example}
\end{figure}

When comparing the S/N in the measured CCs on the between the data (from KiDS and KiDZ) and MICE2, we find that the
noise level in the simulation is typically much smaller in most redshift bins, especially in the sixth tomographic bin
(see top panel of Fig.~\ref{fig:adapt_example}).  The reason for this difference is not entirely clear. Possible reasons
could be differences in the selection functions applied to MICE2 for both the calibration and \kidslegacy\ data, as
compared to the real data. There may also be a difference in the overall clustering amplitude found in the data and the
simulations, especially on small scales. Additionally, the key difference between the data and MICE2 is that the mock
photometry is perfectly uniform such that the effects of variable depth \citep{heydenreich/etal:2020} are not present
in the simulation. Finally, there may be other systematic effects in the data for which our MICE2 data does not account,
for example, related to fibre-collisions in the spectroscopic reference sample, for which DESI (for example) require
special pair-weights \citep{bianchi/etal:2018} that we currently cannot integrate into our correlation estimator.

Since there was no clear explanation for this difference between data and simulation, we opted to adapt the measurements
on the mocks by adding additional Gaussian noise to the values and inflating the error bars such that they match the
data. We computed the right amount of noise required from the difference between the covariances of the
inverse-variance-combined measurements of data and mock.  Figure~\ref{fig:adapt_example} shows a comparison of the
original measurements in MICE2 and how the adapted version compares to the data measurements in the sixth tomographic
bin.

\subsection{Modelling of the measurements}\label{sec:shift_fit}

A long-standing issue within clustering redshift measurements is that the measured correlation functions are, by
definition, not probability densities and must therefore be modelled in some way. A number of different approaches
\citep[e.g.][]{johnson/etal:2017,stoelzner/etal:2021,gatti/etal:22,naidoo/etal:23} have been implemented to mitigate the
frequently arising negative correlation amplitudes, as they represent only a noisy realisation of the underlying
redshift distribution.  Given the dominant sensitivity of cosmic shear to the mean redshifts of the tomographic bins, we
opted for simplicity here and use the SOM \nz\ as a model that we fit to the CC data via a simple shift in redshift,
\dshift{}, and a free normalisation\footnote{ The normalisation is necessary because we cannot expect the data points to
be properly normalised a priori due to galaxy bias or other systematic effects.  }, $A$, such that
\begin{equation}
  n_{\rm model}(z) = A \, n(z - \dshift{}) \, .
  \label{eq:shift}
\end{equation}
This approach, however, can be quite sensitive to single data points with small variance, which can be aggravated in
case of mismatches between the shape of the CCs and the model \nz. This is of particular importance since there seems to
be some additional intrinsic scatter in the data CCs that exceeds the variance that one expects given the uncertainties
of the measurements. This intrinsic scatter is most obvious in bins five and six and, in particular, at high redshifts.

Observational systematics might introduce such erroneous additional correlations and could be reduced in future work by
using organised random catalogues \citep{yan/etal:2024}. Here, we opted for a simpler empirical correction and extended
our fit model with an additive error term $f \, (1+z)$ with free amplitude, $f$, such that the combined uncertainty $s$
for each measurement was
\begin{equation}
  s = \sqrt{\Delta n_{\rm CC}^2 + f^2 \, (1+z)^2} \; .
  \label{eq:errterm}
\end{equation}
We integrated this error term into our likelihood and marginalised over $f$ when determining the shift parameters
\dshift{} and the amplitudes $A$.

This process of measuring and modelling clustering redshifts with a reference \nz\ was tested on the MICE2 mocks, where
we could additionally use the true redshift distribution as a fit model in Eq.~(\ref{eq:shift}). This allowed us to
verify the robustness of, and determine any biases inherent to, our methodology. A key difference with respect to the
data, however, is that the additional intrinsic scatter, parameterised by the $f$-term, is not present in MICE2. We
therefore determined $f$ on the data by fitting each tomographic bin and added the expected intrinsic scatter to the
mock measurements. We perturbed the data points (but not their uncertainties, which were already adapted via the method
described in Sect.~\ref{sec:noise_adapt}) with Gaussian noise with a variance of $f^2 \, (1+z)^2$, which is also
included in the example shown in Fig.~\ref{fig:adapt_example}.  Since we must avoid biases that may arise from a certain
random realisation of the added scatter, we always created 100 realisations, fit each of them independently, and
computed the mean and variance of \dshift{} from all realisations.
 
\section{Simulation results}\label{results: simulations} 

In this section, we present the results of the redshift distribution estimation from our simulated galaxy samples. In
Sect.~\ref{results: SOM} we detail the redshift distributions and bias parameters estimated using our SOM algorithm.
In Sect.~\ref{results: CC} the redshift distributions and bias parameters estimated using the CC method are shown. 
Within these sections, we cover the sensitivity tests that were performed to determine the robustness of the methods. 

\begin{table*}
  \caption{Redshift distribution estimates from simulations.}\label{tab:SOMbiases}
  \resizebox{\textwidth}{!}{
  \begin{tabular}{ccccrrrrrrc}
    \hline\hline
    Simulation  & Run & Desc. & Statistic & \multicolumn{6}{c}{Tomographic Bin} & Described \Tstrut \\
                & ID  &       &           & \multicolumn{1}{c}{1} & \multicolumn{1}{c}{2} & \multicolumn{1}{c}{3} &
                \multicolumn{1}{c}{4} & \multicolumn{1}{c}{5} & \multicolumn{1}{c}{6} & in: \\
    \hline\Tstrut
    \multirow{25}{*}{\skills}
&  \multirow{4}{*}{$\left[{\textbf A}\right]:$} 
     &  \multirow{4}{*}{\shortstack{Fiducial\\(Pvol \&\\ Swgt)}}
     &     $\langle\hat\mu_z\rangle$    &  0.315 & 0.472 & 0.604 & 0.798 &  0.998 &  1.312 &\multirow{4}{*}{\ref{SOM:fiducial}} \\
     &&  & $\langle\delta_{z,0}\rangle$ &  0.026 & 0.038 & 0.006 &-0.005 & -0.095 & -0.196 &  \\
     &&  & $\langle\delta_z\rangle$     & -0.026 & 0.014 &-0.002 & 0.008 & -0.011 & -0.054 &  \\
     &&  & $\sigma_{\delta z}$          &  0.002 & 0.001 & 0.002 & 0.001 &  0.002 &  0.004 &  \\ 
     \cline{2-11}\Tstrut
     &  \multirow{4}{*}{$\left[{\textbf B}\right]:$} 
     &  \multirow{4}{*}{\shortstack{Single Shear\\Realisation}}
     &    $\langle\hat\mu_z\rangle$       &  0.316 &  0.472 &  0.605 &  0.798 &  0.999 &  1.313    &\multirow{3}{*}{\ref{SOM: oneshear}}\\
     && & $\langle\delta_z\rangle$        & -0.025 &  0.013 & -0.001 &  0.008 & -0.012 & -0.053 & \\
     && & $\sigma_{\delta z}$             &  0.002 &  0.001 &  0.002 &  0.001 &  0.004 &  0.004   & \\
     && & $\Delta \langle\delta_z\rangle$ &  0.001 & -0.000 &  0.001 & -0.001 &  0.000 &  0.002 & \\
                                             
     \cline{2-11}\Tstrut
     &  \multirow{1}{*}{$\left[{\textbf C}\right]:$} 
     &  \multirow{1}{*}{\shortstack{Algor. var.}}
                & $\sigma_{\delta z}$   & 0.001  & 0.001 & 0.001 & 0.001 & 0.001 & 0.003 &  \ref{SOM: matching} \\
     \cline{2-11}\Tstrut
                                             
     &  \multirow{4}{*}{$\left[{\textbf D}\right]:$} 
     &  \multirow{4}{*}{\shortstack{$z$-dependant\\Shear}}
     &    $\langle\hat\mu_z\rangle$       &  0.314 &  0.470 &  0.603 &  0.798 &  0.999 &  1.313 &\multirow{3}{*}{\ref{SOM: varshear}}\\
     && & $\langle\delta_z\rangle$        & -0.027 &  0.012 & -0.003 &  0.008 & -0.010 & -0.051 & \\
     && & $\sigma_{\delta z}$             &  0.002 &  0.001 &  0.002 &  0.001 &  0.002 &  0.005 & \\
     && & $\Delta \langle\delta_z\rangle$ &  0.001 &  0.002 &  0.001 &  0.000 & -0.001 & -0.003 & \\
     \cline{2-11}\Tstrut
     &  \multirow{5}{*}{$\left[{\textbf E}\right]:$} 
     &  \multirow{5}{*}{\shortstack{No stellar\\contam.}}
     &     $\langle\hat\mu_z\rangle$    & 0.316 & 0.470 & 0.608 & 0.800 & 1.008 & 1.334 &\multirow{4}{*}{\ref{SOM: no stars}} \\
     &&  & $\langle\delta_{z,0}\rangle$ & 0.041 & 0.040 & 0.011 &-0.003 &-0.091 &-0.171 &                     \\
     &&  & $\langle\delta_z\rangle$     &-0.023 & 0.008 &-0.001 & 0.006 &-0.007 &-0.039 &                \\
     &&  & $\sigma_{\delta z}$          & 0.002 & 0.001 & 0.002 & 0.001 & 0.003 & 0.004 &                     \\
     &&&$\Delta \langle\delta_z\rangle$ &0.0024 &-0.005 & 0.001 &-0.002 & 0.005 & 0.015 &                 \\
     \cline{2-11}\Tstrut
     &  \multirow{4}{*}{$\left[{\textbf F}\right]:$} 
     &  \multirow{4}{*}{\shortstack{No calib.\\weights}}
     &     $\langle\hat\mu_z\rangle$   & 0.346 &  0.497 &  0.637 &  0.803 &  0.982 &  1.302 &\multirow{4}{*}{\ref{SOM: calib weights}}\\
     &&  & $\langle\delta_{z,0}\rangle$&-0.004 &  0.035 &  0.006 & -0.000 & -0.087 & -0.163&                 \\
     &&  & $\langle\delta_z\rangle$    & 0.008 &  0.038 &  0.031 &  0.011 & -0.029 & -0.065 &                 \\
     &&  & $\sigma_{\delta z}$         & 0.003 &  0.000 &  0.002 &  0.000 &  0.002 &  0.005 &                     \\
     \cline{2-11}\Tstrut
     &  \multirow{4}{*}{$\left[{\textbf G}\right]:$} 
     &  \multirow{4}{*}{\shortstack{Swgt only\\\tiny (no prior\\\tiny volume\\\tiny weight)}}
     &     $\langle\hat\mu_z\rangle$    & 0.345 &  0.494 &  0.636 &  0.807 &  0.997 &  1.298 &\multirow{4}{*}{\ref{SOM: shape weights}} \\
     &&  & $\langle\delta_{z,0}\rangle$ &-0.018 &  0.030 & -0.002 & -0.006 & -0.096 & -0.193 &                \\
     &&  & $\langle\delta_z\rangle$     & 0.002 &  0.035 &  0.029 &  0.016 & -0.014 & -0.068 &                 \\
     &&  & $\sigma_{\delta z}$          & 0.003 &  0.001 &  0.002 &  0.001 &  0.003 &  0.005 &                     \\
     \cline{2-11}\Tstrut
     &  \multirow{4}{*}{$\left[{\textbf H}\right]:$} 
     &  \multirow{4}{*}{\shortstack{Pvol only\\\tiny (No calib.\\\tiny shape\\\tiny weight)}}
     &     $\langle\hat\mu_z\rangle$   & 0.318 &  0.472 &  0.608 &  0.801 &  1.008 &  1.335 &\multirow{4}{*}{\ref{SOM: prior weights}}\\
     &&  & $\langle\delta_{z,0}\rangle$& 0.043 &  0.043 &  0.015 &  0.002 & -0.086 & -0.164 &                \\
     &&  & $\langle\delta_z\rangle$    &-0.020 &  0.013 &  0.004 &  0.011 & -0.001 & -0.032 &                \\
     &&  & $\sigma_{\delta z}$         & 0.002 &  0.001 &  0.001 &  0.001 &  0.003 &  0.005 &                     \\
    \hline
    \hline\Tstrut
     \multirow{4}{*}{\shortstack{\skills\\trunc.}}
     &  \multirow{4}{*}{$\left[{\textbf I}\right]:$} 
     &  \multirow{4}{*}{\shortstack{Pvol only\\\tiny (no calib.\\\tiny shape\\\tiny weight)}} 
     &     $\langle\hat\mu_z\rangle$    & 0.320 &  0.469 &  0.604 &  0.798 &  0.983 &  1.110  &\multirow{4}{*}{\ref{SOM: skills_mice2}} \\
     &&  & $\langle\delta_{z,0}\rangle$ & 0.032 &  0.041 &  0.022 &  0.001 & -0.074 & -0.091 & \\ 
     &&  & $\langle\delta_z\rangle$     & 0.001 &  0.015 &  0.017 &  0.014 & -0.004 & -0.014  &\\ 
     &&  & $\sigma_{\delta z}$          & 0.001 &  0.001 &  0.001 &  0.001 &  0.001 &  0.005  &  \\ 
\hline\Tstrut
     \multirow{5}{*}{MICE2} 
     &  \multirow{5}{*}{$\left[{\textbf J}\right]:$} 
     &  \multirow{5}{*}{\shortstack{Pvol only\\\tiny (no calib.\\\tiny shape\\\tiny weight)}}
     &      $\langle\hat\mu_z\rangle$       & 0.304 &  0.468 &  0.593 &  0.796 &  0.965 &  1.097 & \multirow{5}{*}{\ref{SOM: skills_mice2}} \\
     &&   & $\langle\delta_{z,0}\rangle$    & 0.048 &  0.042 &  0.021 &  0.024 & -0.039 & -0.074 &\\
     &&   & $\langle\delta_z\rangle$        & 0.010 &  0.020 &  0.009 &  0.034 &  0.014 & -0.011& \\
     &&   & $\sigma_{\delta z}$             & 0.002 &  0.002 &  0.002 &  0.001 &  0.002 &  0.007 &   \\
     &&   & $\Delta \langle\delta_z\rangle$ & 0.009 &  0.004 & -0.008 &  0.020 &  0.017 &  0.003 & \\
\hline\Tstrut
  \end{tabular}  
}
\tablefoot{
Entries above include abbreviations for prior volume weights (`Pvol'), \lensfit\ shape weights (`Swgt'), 
Statistics in the table include: 
  \begin{tabular}{ll}
    $\langle\hat\mu_z\rangle$: & average of mean redshifts of
    all realisations of the calibration sample\\
    $\langle\delta_{z,0}\rangle$: & average bias of the mean redshift
    prior to SOM\\
    $\langle\delta_{z}\rangle$: & average bias of the mean redshift
    after SOM weighting\\
    $\sigma_{\delta z}$: & uncertainty of $\langle\delta_{z}\rangle$, where values of $\leq 0.01$ are considered negligible.\\
    $\Delta \langle\delta_z\rangle$: & difference in biases (after SOM
                                       weighting) between scenario and
                                       reference, i.e. $\langle\delta_{z}\rangle -
\langle\delta_{z}\rangle_{\rm ref}$
  \end{tabular}
}
\end{table*}

\subsection{SOM redshift distributions}\label{results: SOM} 
Table~\ref{tab:SOMbiases} summarises the results of our SOM redshift distribution calibration using our various
simulations. The table presents multiple statistics, which we outline here.  $\langle\hat\mu_z\rangle$ is the average of
the estimated redshift distribution first moments (i.e. means) under realisations of the calibration sample (see
Sect.~\ref{sec:simulations}). $\langle\delta_{z,0}\rangle$ is the average bias in the first moments prior to any
re-weighting by the SOM, under realisations of the calibration sample, relative to the true weighted redshift
distribution of the lensing sample (which is naturally unknowable for real data). $\langle\delta_{z}\rangle$ is the
average bias in the first moments after re-weighting by the SOM. $\sigma_{\delta z}$ is the standard deviation of the
distribution of biases after re-weighting by the SOM. Finally, $\Delta \langle\delta_z\rangle = \langle\delta_{z}\rangle
- \langle\delta_{z}\rangle_{\rm ref}$ is the difference of the average biases, after re-weighting by the SOM, between a
given scenario and a reference/fiducial scenario. 

\subsubsection{$\left[{\textbf A}\right]:$ Fiducial calibration}\label{SOM:fiducial}
We computed the \nz\ and bias parameters for our fiducial SOM redshift calibration methodology and parameter set
(Table~\ref{tab:SOMparams}), and subsequently compared alternative analyses (such as sensitivity tests) to these fiducial
results. The results of these sensitivity tests were then folded into our analysis in one of two ways. For tests that are
expected to have no impact on the redshift distributions (such as perturbations to ad hoc parameters of the simulation
construction), we folded differences in recovered \nz\ into our systematic error budget. For tests that are expected to
have some systematic (possibly non-negligible) impact on the redshift distributions (such as alternate sample
weighting), we utilised the resulting \nz\ and bias parameters for use in full end-to-end reruns of the \kidslegacy\
cosmology, and presented these results as alternative cosmological constraints in \citet{wright/etal:2025b}. This is
because the differences in biases that arise from the latter type of analyses are not indicative of systematic effects
in the methodology: rather the samples underlying the analyses have become systematically different. 

\begin{figure*}
  \centering
  \includegraphics[width=0.9\textwidth]{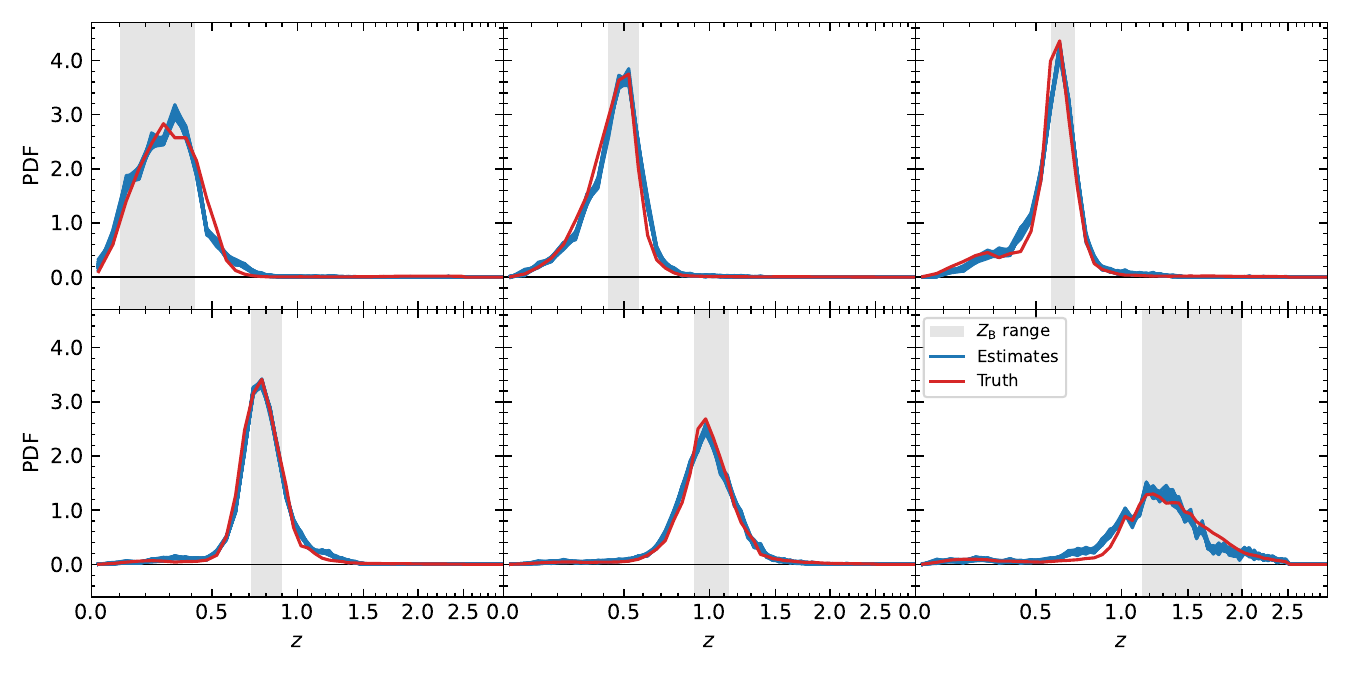}
  \caption{
    Redshift distributions for the fiducial SOM calibration methodology, computed using the \skills\ simulations.
    The ensemble of blue polygons show the \nz\ constructed with different realisations of the calibration sample.
    Redshift distributions include the new gold-weighting for the wide-field sample, and prior volume weighting and
    shape-measurement weights for the calibration samples.
  }
  \label{fig:SOMnz_sims}
\end{figure*}

Resulting \nz\ for our fiducial SOM redshift calibration methodology are provided in Fig.~\ref{fig:SOMnz_sims}.  The
redshift distributions are well constrained to the tomographic bin limits for all six tomographic bins used in
\kidslegacy, with the third tomographic bin showing the largest outlier population (an extended tail to low redshifts)
and the sixth tomographic bin showing a bias to somewhat lower redshifts than targeted by the photo-$z$ cuts. The figure
shows the full range of redshift distributions estimated with realisations of the calibration sample as a filled
polygon. Overlaid is the true target \nz, which is also shown as a polygon but which has, in effect, vanishingly small
area. The figure demonstrates the consistency of the estimated and true redshift distributions: in every bin, our
estimated \nz\ are able to successfully capture the full complexity of the source redshift distributions with accuracy
and precision. 

Our fiducial redshift distributions and biases are somewhat different to those presented in previous \kids\
analyses, such as \citet{wright/etal:2020a}. In particular, the uncertainties on the bias parameters are an order of
magnitude smaller than in previous work, down from $\sigma_{\Delta z} \approx 0.01$ to $\sigma_{\Delta z} \approx
0.001$. This reduction is attributable primarily to the use of our matching algorithm, which forces the calibration
sample redshift distribution to be identical under all realisations (that is, equal to the observed redshift
distribution), regardless of the underlying large-scale structure. These fiducial uncertainties form our base
uncertainty for the SOM calibration method, which are then increased as needed to encompass the systematic uncertainties
determined for the method in the sections below. 

Finally, we note again that, compared to previous SOM redshift calibration work within \kids\ \citep[see
e.g.][]{wright/etal:2020a,hildebrandt/etal:2021,vandenbusch/etal:2022}, the redshift calibration process here
utilises different tomographic binning and a much larger spectroscopic calibration sample. Furthermore, we implement various
weights and selections on the calibration side not previously implemented in \kids. This makes direct comparison between
our fiducial results and results presented in previous \kids\ work difficult; it would be inappropriate, for example, to
apply calibrations presented here to previous work with \kids\ (without redoing the redshift calibration entirely). 

\subsubsection{$\left[{\textbf B}\right]:$ Single shear realisation}\label{SOM: oneshear} 
Our fiducial analysis used the full \skills\ simulation set, including eight realisations of the simulated photometry
catalogues that were generated by applying four uniform shears and two position angle rotations to all sources in the
$108$ \sqdeg\ of simulated sky. These realisations are useful as each has an independent realisation of photometric
noise, allowing us to increase the effective size of our simulated catalogue by a factor of 8. This does not,
however, reduce the sample variance contribution to the analysis, as all photometric realisations are drawn from the
same underlying large-scale structures. 

The motivation for the use of all shear realisations in the fiducial pipeline is a practical one: we require the shear
realisations for shape calibration, and all sources must be appropriately gold-weighted. However, using the multiple
photometric realisations considerably increases the runtime of our calibration. Therefore, in the interest of speed and
reducing unnecessary power consumption, we only utilised a single shear realisation for much of the redshift calibration
testing presented here. 

In that respect, we first verified that the results that we find for our fiducial redshift calibration process are
unchanged under reduction of the number of sources available for calibration testing. These results are presented in
Table~\ref{tab:SOMbiases}: we found that the use of a single shear realisation produces bias estimates that are
consistent with the fiducial case to better than $|\Delta \langle{\delta z}\rangle|=0.0012$ in all bins. Furthermore,
the scatter of the calibrated biases is always consistent between the single and fiducial shear runs, with the
exceptions that the scatter is increased in bin one and decreased in bin six. However, as the absolute size of the
scatter is still relatively small, we conclude that there is unlikely to be a systematic bias introduced in our testing
process by using a single shear realisation in redshift calibration testing. 

\subsubsection{$\left[{\textbf C}\right]:$ Matching algorithm validation}\label{SOM: matching} 
We validated the robustness of our calibration to the ad hoc matching algorithm parameters (described in
Sect.~\ref{sec:sim_match}) by constructing multiple realisations of the calibration sample using perturbations on the
fiducial choices.  We then propagated these modified calibration samples through the redshift calibration pipeline and
compared the biases that we estimated to those from the fiducial setup.  For perturbations in the matching parameters,
we tested three choices of redshift window size ($1000$, $2000$, and $10\,000$ sources, respectively), 10 perturbations
to the matching algorithm feature space (dropping one band at a time), matching on colours and a reference magnitude,
and matching without \photoz\ information.  We found that the matching algorithm is extremely robust to each of these
perturbations, with typical changes in the bias (from that measured in the fiducial case) of order $\Delta
\langle{\delta z}\rangle=10^{-4}$. For use in our uncertainty budget, we computed the standard deviation of the
recovered bias parameters under the different realisations of the matching algorithm; these are reported in
Table~\ref{tab:SOMbiases}. In all bins the scatter introduced in the bias from the matching algorithm perturbations is
less than half the scatter in the spatial/photometric realisations of the calibrations samples in the fiducial case
(i.e. $\sigma_{\Delta \langle{\delta z}\rangle}<0.5\sigma_{\langle{\delta z}\rangle,{\rm fid}}$). The maximal
uncertainty in the bias introduced by our matching algorithm perturbations is $\sigma_{\Delta \langle{\delta
z}\rangle}=0.0029$, in the sixth tomographic bin. 

\subsubsection{$\left[{\textbf D}\right]:$ Impact of redshift-dependent shear}\label{SOM: varshear} 
Recent analyses by \citet{maccran/etal:2022} highlighted the interplay between redshift distribution estimation and
shape-measurement calibration for ongoing stage-III cosmic shear surveys. This was a primary motivating factor for the
construction of the \skills\ simulations by \citet{li/etal:2022}, who used the redshift-dependent shear realisations of
\skills\ to estimate the impact of higher-order blending effects on an analysis such as \kidslegacy. The found a
measurable, but otherwise minor, impact of redshift-dependent shear and blending on the computation of shape-measurement
bias in \kidslegacy. We repeated this analysis in the context of redshift distribution bias estimation. 

In the context of redshift distribution estimation, redshift-dependent shear primarily manifests as a modification to
the weights of sources, which are blended across significant redshift baselines. This is therefore a fairly minor
perturbation to the properties used to estimate our redshift distributions and, as a result, the impact that this effect
has on our recovered \nz\ is similarly minor. In all bins, the measured change in bias, which we find from analysing
\skills\ simulations with redshift-dependent shear, is $\Delta \langle\delta_z\rangle < 0.003$. Therefore, we conclude
that redshift-dependent shear in blends is a subdominant source of bias, for both shape-calibration and
redshift-distribution calibration, at the sensitivity of \kidslegacy.   

\subsubsection{$\left[{\textbf E}\right]:$ Impact of stellar contamination}\label{SOM: no stars} 
A primary development of the \skills\ simulation over those previously used in \kids\ redshift calibration is the
full-complexity inclusion of image-based source extraction and modelling. An outcome of this process is that the
simulation no longer includes an artificially perfect stellar rejection. \cite{li/etal:2022} demonstrate using their
\konek\ simulation that the \kids\ lensing sample is contaminated by a residual population of stars, after all
selections/cleaning, at the level of $\sim0.56\,\%$. These sources contaminate the shear measurements of the survey, and
are required to be calibrated-out using these simulations (through correction of additive and multiplicative shear
measurement biases). 

An additional effect not yet analysed in \kids, however, is the effect that these sources have on the redshift
distribution bias estimates. Stellar contamination influences the redshift distribution as a population of sources at
$z=0$, which (depending on the amount of contamination) can possibly contribute non-negligibly to the location of the
distribution mean. Additionally, stellar sources are not represented in the spectroscopic calibration sample, as
spectroscopic surveys are generally able to reliably flag and remove stars. This means that stars will act to
erroneously boost the significance of redshifts in the reconstructed \nz, where they coincide with galaxy colours. 

We investigated the significance of the contribution of stars to the redshift distributions from \skills\ by running our
calibration pipeline assuming perfect stellar rejection, and comparing the resulting \nz\ distributions and biases to
those from our fiducial run. These results are presented in Table~\ref{tab:SOMbiases}. We find that the difference in
bias with and without the stellar contamination is of similar order as the uncertainty in the bias estimates between
realisations: the maximal difference in the recovered bias is $|\Delta\langle\delta_z\rangle|=0.0078$ in the sixth
tomographic bin, which is a roughly $2\sigma$ deviation from the fiducial bias uncertainty (which itself is a lot
smaller than the conservative final uncertainty; see Sect.~\ref{SOM: skills_mice2}). 

\subsubsection{$\left[{\textbf F}\right]:$ Calibration-side weighting}\label{SOM: calib weights} 
In previous \kids\ analyses, redshift calibration has always been performed without weights utilised on the calibration
side of the SOM/direct calibration process. This is primarily because the spectroscopic calibration fields were (and
largely remain to be) not wide-field shear fields. However, as discussed in Sect.~\ref{sec:SOM}, in \kidslegacy\ we
implement additional weighting on the calibration side in an effort to mitigate systematic biases: shape-measurement
weights, and prior-volume weights. We tested the impact of removing these weights one at a time in Sects.~\ref{SOM: shape
weights} and \ref{SOM: prior weights}. For closer compatibility with previous work, however, we also tested the redshift
calibration process when not including either of these weights. The results are presented in Table~\ref{tab:SOMbiases}. 

In terms of bias, we found that the inclusion of the weighting (i.e. Run ID [\textbf{A}]) produces smaller biases in some
bins than the unweighted scenario (i.e. Run ID [\textbf{E}]). These shifts are partly larger than the uncertainty
estimates. Bins two, four, and five see reductions in the absolute value of the bias (going from [\textbf{E}] to
[\textbf{A}]) at the $2-3\sigma$ level. However, the opposite effect is true in some other bins: bins one and three see
a $2\sigma$ increase in absolute value of the bias when including the calibration side weights. Of note is the mechanism
for bias to change on the redshift distributions without corresponding change in the estimated redshift distribution
means. The weighting on the calibration side influences the effective redshift distribution of the wide-field sample
(i.e. the truth) through the gold-weight, which itself is modified by the relative up- and down-weighting of calibration
sources by the shape and prior volume weights. 

Furthermore, the bias in the redshift distribution means prior to SOM weighting is considerably larger in bins two to
four without the calibration-side weighting. This indicates that the weighting applied in the fiducial case brings the
calibration and wide-field samples closer together prior to the re-balancing of the colour space via the SOM weights. 

Overall, the weights on the calibration side were found to have a non-negligible impact on the redshift distribution
biases. The individual impact of the shape-measurement weights and prior-volume weights are discussed in the following
sections. Given the changing nature of the wide-field sample \nz\ under the calibration-side weighting, we opted
not to include these changes in bias as a systematic component in our cosmological analysis. Rather we instead reserved
the two sets of redshift distributions and weighted source samples for separate cosmological analyses. While the redshift
distributions are not directly comparable and the analysis presented here does not strongly favour one approach, the
cosmological parameters should still be highly consistent for both scenarios. In practice, we use the maximally realistic 
setup for our fiducial case: including both prior volume weights and shape weights on the calibration side. 

\subsubsection{$\left[{\textbf G}\right]:$ Impact of shape measurements}\label{SOM: shape weights} 
We next tested the influence on the redshift distributions biases when implementing weighting on the calibration side
using only shape-measurement weights. We compared these shape-weight-only biases to those from both the fiducial and
no-calibration-weighting results presented in Table~\ref{tab:SOMbiases}. 

We first note that the application of the shape-measurement weights to the calibration sample initially has the
counter-intuitive effect of dramatically increasing the bias in the redshift distributions before SOM weighting. 
This is particularly clear in the fifth and sixth tomographic bins; in bin six, the bias exceeds $\delta_z=0.1$ prior to
SOM weighting. This suggests that the application of the shape-measurement weights alone acts to increase the disparity
between the calibration and wide-field datasets at high redshift. We see the inverse effect, however, in bins two-four,
suggesting that there the inclusion of shape weights makes the calibration sample more representative of the wide-field
data. 

After SOM weighting, however, the situation is clearer: the redshift distribution biases are
significantly improved in all bins except for the first, where the calibration side shape-measurement weights are not
able to correct the over-correction of the SOM weighting (discussed in Sect.~\ref{SOM: calib weights}). Overall, the
addition of calibration side shape weights produces a reduction in bias compared to the unweighted result, particularly
in the bins most sensitive to cosmic shear. 

\subsubsection{$\left[{\textbf H}\right]:$ Impact of prior weights}\label{SOM: prior weights} 
The inclusion of prior volume weights on the calibration side acts to change the relative importance of calibration
sources that reside in cells containing colour-redshift degeneracies. The results when computing redshift distributions
using only calibration-side prior volume weights are shown in Table~\ref{tab:SOMbiases}. Firstly, it is clear that the
inclusion of the prior volume weights brings the redshift distributions before SOM weighting much closer to those of the
wide-field calibration sample: biases in bins two to six all reduce by $|\Delta \langle\delta_{z,0}\rangle| \in[0.01,0.03]$. 
This indicates that the prior volume weights are having a positive effect in removing the systematic differences between
the calibration and wide-field data along the redshift-axis. 

After SOM weighting, we again found biases that were improved in some bins and degraded in others: in the higher redshift
tomographic bins we see consistent further reduction in biases after the SOM weighting; however, we again see
over-correction in the lower redshift bins. Relative to the results when using shape-only weighting we see that the
prior-volume-only weighting produces slightly poorer bias recovery, but that the bias is reduced relative to the
implementation without any calibration-side weighting in the majority of bins. 

\subsubsection{$\left[{\textbf{I-J}}\right]:$ \skills\ versus MICE2}\label{SOM: skills_mice2} 
We verified the computation of biases using two different simulated datasets, as a means of estimating the robustness of
our calibration procedure to the assumptions inherent to a single simulation. To do this, we applied our \nz\ estimation
algorithm to samples constructed in the \skills\ (Sect.~\ref{sec:SKILLS}) and MICE2 (Sect.~\ref{sec:MICE2}) datasets,
where the only a priori modification to the simulations is to ensure that both cover the same underlying redshift
baseline: $0.07 < z_{\rm true} < 1.42$.  This ensures that this test probes the difference that is attributable to the
construction of the simulation for a consistent population of source galaxies, rather than probing differences in bias
generated by different samples with different redshift extent. Additionally, we opted to compute the difference between
the recovered biases in the regime where we do not include shape measurements on the calibration side of the
computation, as the shape-measurement weights are systematically different between the simulations: MICE2 shape
measurement weights are synthetic and determined by matching colours to the observed data, rather than being measured
from images as in \skills. 

Resulting redshift distribution biases for our \skills\ and MICE2 datasets are presented in Table~\ref{tab:SOMbiases}.
For our analysis without calibration side weights and only including the prior volume weights we
found consistent redshift distribution bias parameters for the two simulations at the level of $|\Delta \delta_z|\la
0.01$. This indicates that there is an inherent uncertainty floor in the accuracy to which we can estimate the redshift
distribution bias parameters from our simulations, driven by the realism of the simulations themselves. Such an error
floor is in reality somewhat conservative, as the MICE2 simulations here are known to lack realism in many regards (not
the least of which is the lack of imaging and the use of purely analytic photometric noise realisations). Thus, it is
expected that these simulations ought to diverge in their realism, with \skills\ being the more trustworthy reference.
Nonetheless, we opted to utilise this $|\Delta \delta_z|\la 0.01$ systematic difference in our computation of the redshift
distribution bias priors, by implementing this as an uncertainty floor in the prior specification. 

\subsection{CC redshift distributions}\label{results: CC}
Similar to the SOM analysis, we first tested our calibration methodology on MICE2. Since we needed to measure two-point
statistics on the simulation, we required a slightly different version from those reported in the section above; the
version used here applies the matching algorithm neither on the calibration sample, nor the photometric data to ensure
that the clustering properties of the resulting gold sample are preserved. In addition, we found that, when comparing the
shift-fit values \dshift{} (Eq.~\ref{eq:shift}) to the SOM bias (Eq.~\ref{eqn:som_bias}), it was preferential to compute
the SOM bias as the difference in the median of the redshift distribution instead of the mean, by defining
\begin{equation}
  \delta z_{\rm med} = \text{med}[\somnz] - \text{med}[\truenz] \, .
\end{equation}
The reason is that the shift-fitting is not sensitive to any outlier populations at the tails of the redshift
distribution, which are reflected in the mean, but not the median of the distribution. These values are listed in column
2 of Table~\ref{tab:CCshifts} and are computed from a single shear realisation.

\begin{table}
  \caption{
    Different redshift bias estimates per tomographic bin obtained from the ensemble of MICE2 CC realisations.
  }
  \label{tab:CCshifts}
  \centering
  \renewcommand{\arraystretch}{1.2}
  \begin{tabular}{clll}
\hline\hline
Bin & \multicolumn{1}{c}{$\delta z_{\rm med}$} & \multicolumn{1}{c}{$\langle\dshift{SOM}\rangle$} & \multicolumn{1}{c}{$\delta z_{\rm med} - \langle\dshift{SOM}\rangle$} \Tstrut\Bstrut \\
\hline
1 & $\phantom{-}0.019 \pm 0.010$ & $\phantom{-}0.014 \pm 0.007$ & $\phantom{-}0.005 \pm 0.012$ \Tstrut \\
2 & $\phantom{-}0.058 \pm 0.010$ & $\phantom{-}0.060 \pm 0.006$ & $-0.002 \pm 0.012$ \\
3 & $\phantom{-}0.049 \pm 0.010$ & $\phantom{-}0.056 \pm 0.006$ & $-0.007 \pm 0.011$ \\
4 & $\phantom{-}0.018 \pm 0.010$ & $\phantom{-}0.015 \pm 0.006$ & $\phantom{-}0.003 \pm 0.012$ \\
5 & $-0.004 \pm 0.010$ & $-0.016 \pm 0.019$ & $\phantom{-}0.012 \pm 0.022$ \\
6 & $-0.022 \pm 0.012$ & $-0.085 \pm 0.071$ & $\phantom{-}0.063 \pm 0.072$ \Bstrut \\
\hline
\end{tabular}   \tablefoot{
    The values listed here are the median SOM bias, the mean and standard deviation of the shift-fit parameter obtained by fitting
    the CCs with the true redshift distribution, followed by fitting with the \somnz, and finally the difference between the SOM
    bias and the corresponding shift-fit parameter. 
    The results in this table are not directly
    comparable to Table~\ref{tab:SOMbiases} as the source samples are inherently different. $\langle\dshift{SOM}\rangle$ is
    duplicated from Table~\ref{tab:CCfits} for comparison.
  }
\end{table}

\subsubsection{CC measurements}
\begin{figure*}
  \centering
  \includegraphics[width=0.9\textwidth]{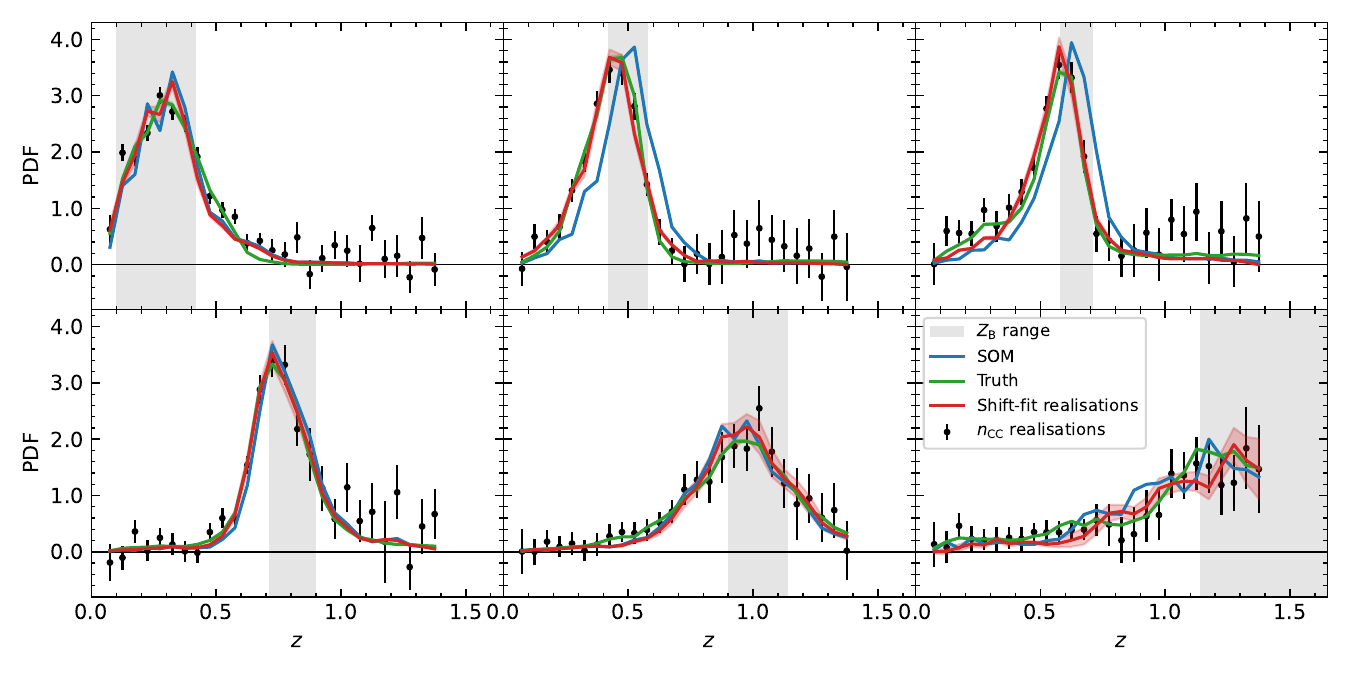}
  \caption{Comparison of the ensemble of MICE2 realisations and their best-fit solutions when using the SOM redshift distributions as
    model for the shift-fit. The black data points indicate the mean and standard deviation (encapsulating the added intrinsic
    scatter) of the CC measurement realisations, the green and blue lines represent the true and SOM redshift distributions,
    respectively. The red line and shaded area (median and 68~\% confidence interval) is the \somnz\ after applying the shift-fit
    parameter value in each realisation.
  }
  \label{fig:nzs_cc_mock}
\end{figure*}

As described in Sect.~\ref{sec:shift_fit}, we did not use the clustering redshift distributions that we measured
directly from MICE2.  Instead, we first applied the noise adaptation scheme and added the level of intrinsic scatter
that we found in the data by fitting the $f$-term (Eq.~\ref{eq:errterm}) on the data and created 100 realisations of the
intrinsic scatter. The mean and scatter of these realisations are shown in Fig.~\ref{fig:nzs_cc_mock}. They closely
trace the underlying true redshift distribution of the simulated \kidslegacy\ data (green line).

\subsubsection{CCs fitted with true redshift distributions}\label{sec:CC_true_nz}
First, we needed to verify that our updated measurement process and fitting procedure were able to accurately reproduce the true
redshifts of the MICE2 galaxies. Therefore, we fit the CC measurements with the true redshift distribution to check whether the
resulting shift parameter \dshift{true} was consistent with zero in all bins. We compared a number of different correlation measurement
scales and found that excluding the smallest scales ($r < 0.5~\mathrm{Mpc}$) resulted in the least biased shift parameters,
whereas setting the upper limit to $1.5~\mathrm{Mpc}$ still maintained a good S/N in the correlation amplitude.
Therefore, we chose $0.5 < r \leq 1.5~\mathrm{Mpc}$ as fiducial measurement scale for \kidslegacy.

The ensemble of parameter values from all realisations of this setup, fitted with the true redshift distribution, are
listed in Table~\ref{tab:CCfits} and shown by the green data points in Fig.~\ref{fig:fit_shifts}. Most importantly, all
shift-fit values are consistent with zero and $|\langle \dshift{true} \rangle| \, < 0.01$ (except for bin six),
demonstrating that our new methodology is able to produce unbiased estimates of the underlying true redshift
distribution.  In the first four tomographic bins, where most of the redshift distributions are covered by the wide area
spectroscopic samples of 2dFLenS, BOSS, and GAMA, the scatter in \dshift{true} is $0.005$. In bins five and six, which
are dominated by the CCs from DESI, the scatter is significantly larger ($0.017$ and $0.044$) with the largest shift of
$\langle \dshift{true} \rangle = -0.015$ in bin six.

We found that our fitting approach, with the additional error term that accounted for intrinsic scatter, was able (on
average) to reproduce the input value from the measurements of the data. The scatter of the $f$-values between
realisations was similar to the uncertainty of the input value. The distribution of goodness-of-fit values was
consistent with $\chi^2/{\rm dof} = 1$ in all bins, as was expected given the model's ability to increase the magnitude
of the uncertainty.

\begin{table*}
  \caption{
    Parameters and goodness of fit obtained from CC analyses. 
  }
  \label{tab:CCfits}
  \resizebox{\textwidth}{!}{
    \centering
    \renewcommand{\arraystretch}{1.2}
    \begin{tabular}{llcllllll}
\hline\hline
\multicolumn{1}{c}{Dataset} & \multicolumn{1}{c}{Fit model} & & \multicolumn{6}{c}{Tomographic Bin} \Tstrut \\
 & & & \multicolumn{1}{c}{1} & \multicolumn{1}{c}{2} & \multicolumn{1}{c}{3} & \multicolumn{1}{c}{4} & \multicolumn{1}{c}{5} & \multicolumn{1}{c}{6} \Bstrut \\
\hline
\multirow{3}{*}{Data} & \multirow{3}{*}{SOM $n(z)$\,} & \dshift{} & $\phantom{-}0.000 \pm 0.007$ & $\phantom{-}0.028 \pm 0.006$ & $\phantom{-}0.021 \pm 0.006$ & $\phantom{-}0.035 \pm 0.010$ & $-0.018 \pm 0.028$ & $-0.145 \pm 0.091$ \Tstrut \\
 &  & $f$ & $\phantom{-0}0.06 \pm 0.03$ & $\phantom{-0}0.10 \pm 0.03$ & $\phantom{-0}0.08 \pm 0.03$ & $\phantom{-0}0.06 \pm 0.03$ & $\phantom{-0}0.07 \pm 0.03$ & $\phantom{-0}0.07 \pm 0.04$  \\
 &  & $\chi^2_{\rm dof}$ & $\phantom{-0}1.44$ & $\phantom{-0}1.04$ & $\phantom{-0}1.11$ & $\phantom{-0}1.03$ & $\phantom{-0}1.04$ & $\phantom{-0}1.14$ \Bstrut \\
\cline{1-9}
\multirow{6}{*}{\parbox{18mm}{MICE2\\realisations}} & \multirow{3}{*}{True $n(z)$} & $\langle \dshift{} \rangle$ & $\phantom{-}0.003 \pm 0.005$ & $\phantom{-}0.002 \pm 0.005$ & $\phantom{-}0.005 \pm 0.005$ & $-0.004 \pm 0.007$ & $-0.008 \pm 0.017$ & $-0.015 \pm 0.044$ \Tstrut \\
 &  & $\langle f \rangle$ & $\phantom{-0}0.10 \pm 0.02$ & $\phantom{-0}0.12 \pm 0.04$ & $\phantom{-0}0.11 \pm 0.03$ & $\phantom{-0}0.09 \pm 0.03$ & $\phantom{-0}0.07 \pm 0.03$ & $\phantom{-0}0.09 \pm 0.03$  \\
 &  & $\langle \chi^2_{\rm dof} \rangle$ & $\phantom{-0}0.99 \pm 0.08$ & $\phantom{-0}1.01 \pm 0.09$ & $\phantom{-0}1.02 \pm 0.10$ & $\phantom{-0}1.04 \pm 0.12$ & $\phantom{-0}0.96 \pm 0.14$ & $\phantom{-0}0.98 \pm 0.17$ \Bstrut \\
\cline{2-9}
 & \multirow{3}{*}{SOM $n(z)$} & $\langle \dshift{} \rangle$ & $\phantom{-}0.014 \pm 0.007$ & $\phantom{-}0.060 \pm 0.006$ & $\phantom{-}0.056 \pm 0.006$ & $\phantom{-}0.015 \pm 0.006$ & $-0.016 \pm 0.019$ & $-0.085 \pm 0.071$ \Tstrut \\
 &  & $\langle f \rangle$ & $\phantom{-0}0.15 \pm 0.02$ & $\phantom{-0}0.13 \pm 0.04$ & $\phantom{-0}0.15 \pm 0.03$ & $\phantom{-0}0.10 \pm 0.03$ & $\phantom{-0}0.09 \pm 0.03$ & $\phantom{-0}0.11 \pm 0.04$  \\
 &  & $\langle \chi^2_{\rm dof} \rangle$ & $\phantom{-0}0.97 \pm 0.06$ & $\phantom{-0}0.98 \pm 0.08$ & $\phantom{-0}0.97 \pm 0.07$ & $\phantom{-0}1.01 \pm 0.10$ & $\phantom{-0}0.97 \pm 0.10$ & $\phantom{-0}1.05 \pm 0.15$ \Bstrut \\
\hline
\end{tabular}   }
  \tablefoot{
    Results were computed from shift-fits on the data and the 100 MICE2 realisations with
    noise adaptation. The uncertainties quoted for MICE2 refer to the standard deviation of the realisations, not the error of the
    mean.
    \begin{tabular}{ll}
      $D_z$: & $\nz$ shift parameter (see Eq.~\ref{eq:shift})\\
      $f$: & additive error term (see Eq.~\ref{eq:errterm})
    \end{tabular}
  }
\end{table*}

\subsubsection{CCs fitted with SOM redshift distributions}\label{sec:mock_cc_results_wsom}
The second step of verifying our pipeline was fitting the mock CCs with the SOM redshift distributions and comparing the shift
parameters to the bias in the median SOM redshift ($\delta z_{\rm med}$). Ideally, both values should be in agreement if the
difference between the CCs and \somnz\ can, to first order, be rectified by a simple shift in redshift.

\begin{figure}
  \centering
  \includegraphics[width=\columnwidth]{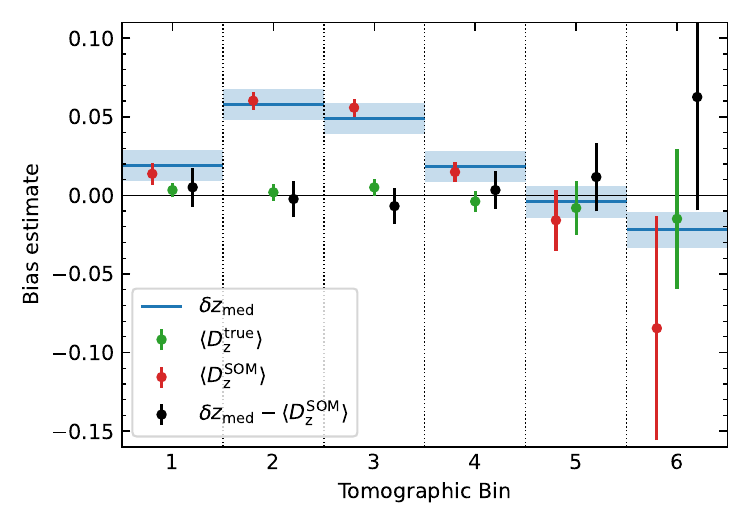}
  \caption{Mean and scatter of the shift parameters obtained from the 100 MICE2 realisations with noise adaptation.
    The blue line and shaded area indicate the bias in the median SOM redshift, the green and red data points indicate the
    mean and scatter of the shift-fit parameters \dshift{} when fitting the realisations with either the true or SOM redshift
    distribution. The black data points represent the difference between the empirical SOM bias estimate and the shift-fit parameter.
  }
  \label{fig:fit_shifts}
\end{figure}

The goodness of fit for these fits is, similar to the case of fitting with the true redshift distributions above,
consistent with $\chi^2_{\rm dof} = 1$.  The fitted $f$-values are up to $50\,\%$ larger compared to the previous case
but are otherwise, with the exception of bin one, fully consistent with the input values from the data when considering the
uncertainty and the scatter between the realisations (see Table~\ref{tab:CCfits}). The shift-parameter values are much
larger when using the \somnz\ as fit model and only consistent with zero in bin five; see the red data points in
Fig.~\ref{fig:fit_shifts}. In bins one to four, $\langle\dshift{SOM}\rangle$ is largely positive, indicating that the
SOM overestimates the median redshift by about $0.015$ in bins one and four and up to $0.060$ in bin two. In general,
the scatter in \dshift{SOM} is similar to the ones obtained from the fits with the true redshift distribution.

While these values indicated a large bias in the SOM redshifts, they are perfectly consistent with the SOM bias reported
for this specific version of MICE2 (compare Table~\ref{tab:CCshifts} and the blue confidence regions in
Fig.~\ref{fig:fit_shifts}). When we take the difference $\delta z_{\rm med} - \langle\dshift{SOM}\rangle$ of the two
bias estimates, the value is consistent with zero considering the scatter of \dshift{SOM} between the noise
realisations, as indicated by the black data points in Fig.~\ref{fig:fit_shifts}. The scatter is about $0.012$ in the
first four bins and $0.022$ and $0.072$ in bins five and six, respectively. The amplitude of the difference is less than
$0.01$ in all but the last two tomographic bins, of which bin six is particularly poorly constrained.

Finally, we confirmed visually in Fig.~\ref{fig:nzs_cc_mock} that applying the shift to the \somnz\ on MICE2 results in a much
better match with the CCs than the unshifted redshifts. Especially in bins two and three, where $\langle\dshift{SOM}\rangle$ is
large, the peak of the distribution closely matches the realisations of the CCs after shifting. There are some remaining residual
differences near the tails of the distributions, which probably explain the increase in the best-fit $f$-values.

\section{Data results}\label{results: data} 

\subsection{SOM redshift distributions} 
\begin{figure*}
    \centering
    \includegraphics[width=0.9\textwidth]{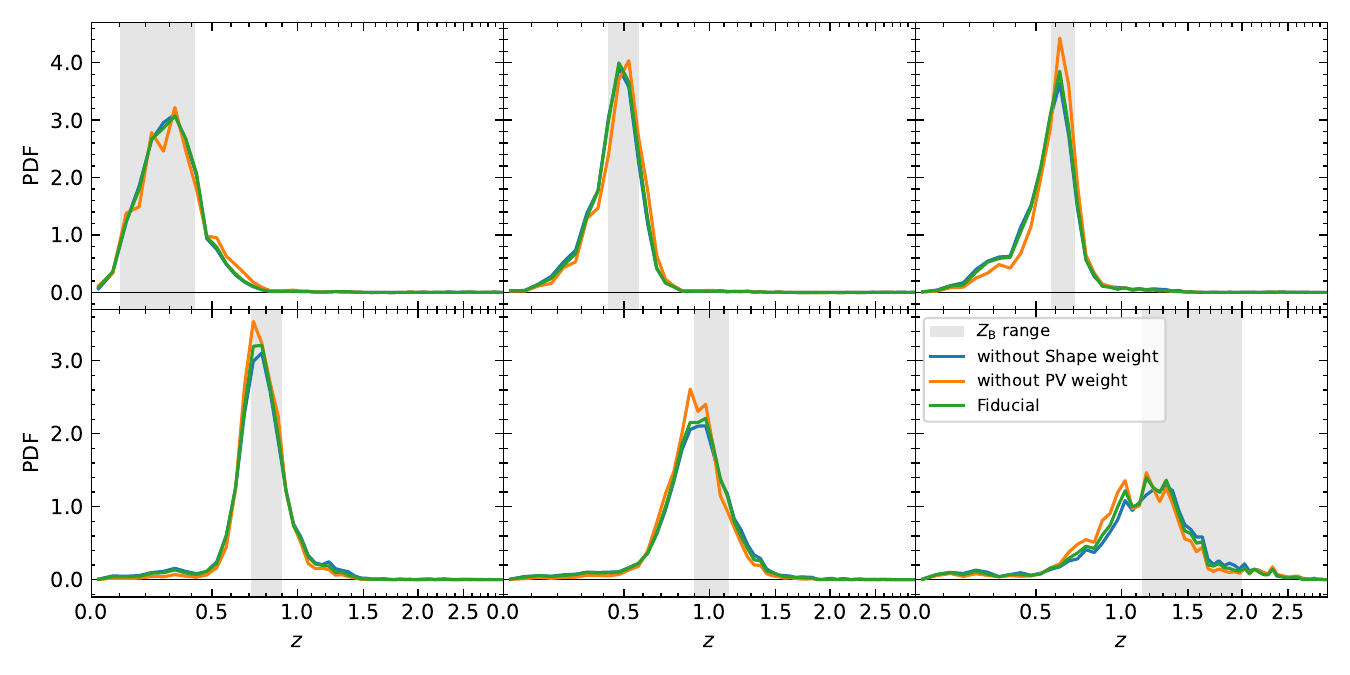}
    \caption{Estimated redshift distributions from the SOM method, for different analysis choices.
      Results are highly consistent, except
      when switching between use or non-use of the prior volume weighting. However, as described in Sect.~\ref{sec:priorweight},
      this difference is shown to be reflected in an increased bias for the non-prior-weighted distributions.
    }
    \label{fig:SOMnz_data}
\end{figure*}

Figure~\ref{fig:SOMnz_data} presents the redshift distributions estimated for the six tomographic bins using the SOM
algorithm, for a range of different analysis choices. The results are largely indistinguishable from one-another, except
that there are two sets of lines; those using the prior volume weighting and those without. As shown in
Sect.~\ref{sec:priorweight}, this is for two reasons. Firstly the prior volume weight acts to smooth the large-scale
structure that is imprinted on the redshift distribution of the calibration sample, leading to smoother estimates of the
wide-field \nz. Secondly, as shown in Fig.~\ref{fig:specz_priorwgt_tomo}, the prior volume weights are able to introduce
systematic shifts in the tomographic redshift distributions of the calibration sample before SOM re-weighting. The
shifts in Fig.~\ref{fig:SOMnz_data} are less significant than the example shown in Fig.~\ref{fig:specz_priorwgt_tomo},
however, demonstrating that the SOM re-weighting has acted to undo some of the prior weight shift. 

The resulting \nz\ in Fig.~\ref{fig:SOMnz_data} are each associated with an estimated bias, and it is worth noting that
the difference in the redshift distributions is almost perfectly compensated by the change in bias predicted by \skills.
Said differently, the response of the data \nz\ to the prior weight is exactly the same as the response of the simulated
\nz\ to the prior weight. This is a further indication that the simulated analysis used to estimate the bias parameters
faithfully reproduces the complexity of the real data. 

Finally, we note the similarity between the redshift distributions estimated on the data and on the simulations. In
particular bins two to five are essentially identical as estimated on \skills\ and on the data. Bin six shows the most
significant differences: the simulated redshift distribution has a considerable smoothly decreasing tail extending to
redshift two, whereas the data \nz\ all truncate fairly abruptly at $z\approx 1.6$. We have not explored the origin of
this difference here, but note that such a difference will be enhanced by slight differences in the signal-to-noise and
size properties of the most distant sources. We attempted to remove such differences using our matching process;
however, this difference may indicate that there is still some residual difference between the data and mock galaxies at
high redshift. 

\subsection{CC redshift distributions}\label{results:CCdata}
Similar to the mock analysis in Sect.~\ref{sec:mock_cc_results_wsom}, we measured the clustering redshifts of the
\kidslegacy\ lensing sample, computed the joint, inverse-variance weighted CC estimate, and performed the shift-fitting
with the fiducial SOM redshift distributions (see above). The final CC measurements are presented in
Fig.~\ref{fig:nzs_cc_data}. In general, these are (in part due to the noise adaptation) very similar to the ensemble
average of the MICE2 realisations (Fig.~\ref{fig:nzs_cc_mock}).

\subsubsection{CCs fitted with SOM redshifts}
\begin{figure*}
  \centering
  \includegraphics[width=0.9\textwidth]{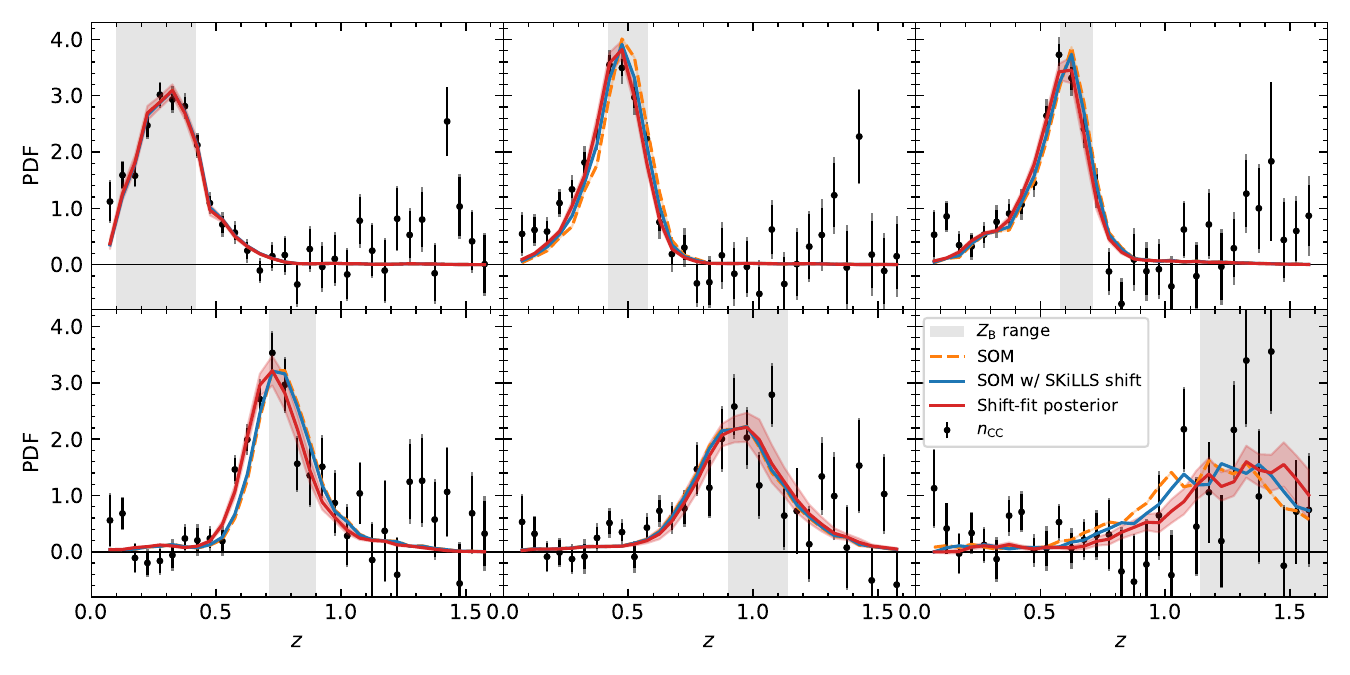}
  \caption{Comparison of the fiducial data CCs, SOM redshift distributions (used as fit model, re-normalised to the fitted amplitude of
    the CCs, and additionally with the estimated SOM bias corrected), and shift-fit posterior median and 68~\% confidence interval.
    The additional grey error bar whiskers indicate the total confidence interval that includes the fitted intrinsic scatter.
  }
  \label{fig:nzs_cc_data}
\end{figure*}

As a result, the shift-fit parameters \dshift{SOM} follow as similar trend as those of MICE2 (Table~\ref{tab:CCfits}).
One major difference is that we observed some additional intrinsic scatter, especially at $z > 1.0$, where DESI dominates
the joint CC measurements, which our simple $f$-term model was unable to fully capture. The best-fit $f$-term is a small
fraction of the total uncertainty of the CC measurements\footnote{When including smaller measurement scales, the
$f$-term contribution becomes significant.} (Fig.~\ref{fig:nzs_cc_data}).

This difference was also reflected in the increased uncertainty of \dshift{SOM} in the last three tomographic bins, where
it reached $\sigma(\dshift{SOM})=0.091$ in bin six. Since we are limited to the maximum redshift of our DESI sample at
$z \approx 1.6$, the consequence is that the mean redshift of bin six was not very well constrained by the clustering
redshifts.  In the five other bins, the uncertainty was found to be at a similar level as MICE2. The magnitude of the
shifts is in general smaller for the first three tomographic bins, reaching a maximum of $0.028$ in bin two, which is
clearly in agreement with visible offset between the SOM and the CCs in Fig.~\ref{fig:nzs_cc_data}. In bins four and
five, the magnitude of the shift is similar to bins two and three, but in bin six it reaches $\dshift{SOM} = -0.145$,
preferring a shift of the \somnz\ at low significance to higher redshifts. Overall, only bins one and five are unbiased
according to the shift-fit after taking the uncertainties into account.

\subsubsection{Comparison to SOM bias from \skills}
\begin{figure}
  \centering
  \includegraphics[width=\columnwidth]{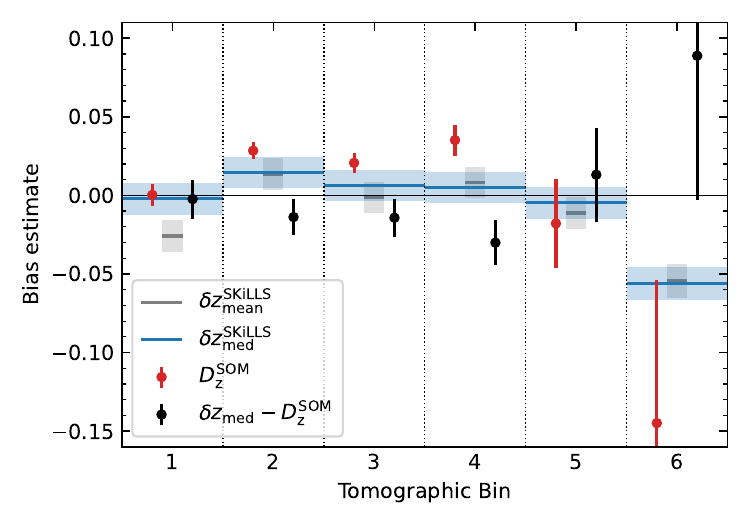}
  \caption{Comparison of the shift parameters obtained from the \kidslegacy\ data and the SOM bias obtained from \skills.
    The short grey and blue lines indicate the bias in the mean and median SOM redshift, the red data points the shift-fit
    parameters \dshift{} when fitting CC measurements with the \somnz. The black data points represent the difference between the
    empirical median SOM bias estimate and the shift-fit parameter.
  }
  \label{fig:fit_shifts_data}
\end{figure}

As a final test, we used the fitted shift-parameters \dshift{SOM} to test how well the SOM calibration of the \skills\ simulation
represented the SOM calibration of the \kidslegacy\ data. Similar to our analysis of the MICE2 data
(Sect.~\ref{sec:mock_cc_results_wsom}) we compared \dshift{SOM} to the bias of the median SOM redshifts obtained from
\skills\ (see Table~\ref{tab:CCshifts_data} and Fig.~\ref{fig:fit_shifts_data}) and found that they are not identical,
but closely follow the same trends. The bias is consistent with zero in bin one, positive for bins two to four, and
smoothly transitions to negative values in bins five and six. In general, the biases indicated by the \dshift{SOM} are
somewhat larger than the SOM biases in \skills. However, when factoring in the uncertainties, the difference $\delta
z_{\rm med} - \dshift{SOM}$ is consistent with zero in all bins, except for bin four, which exhibits an approximately
$2\sigma$ difference, again driven by the CCs preferring larger measured biases. 

\begin{table}
  \caption{Different redshift bias estimates per tomographic bin obtained from the \kidslegacy\ data.
  }
  \label{tab:CCshifts_data}
  \centering
  \resizebox{\columnwidth}{!}{
    \renewcommand{\arraystretch}{1.2}
    \begin{tabular}{clll}
\hline\hline
Bin & \multicolumn{1}{c}{$\delta z_{\rm mean}^{\rm SKiLLS}$ / $\delta z_{\rm med}^{\rm SKiLLS}$} & \multicolumn{1}{c}{$\dshift{SOM}$} & \multicolumn{1}{c}{$\delta z_{\rm med}^{\rm SKiLLS} - \dshift{SOM}$} \Tstrut\Bstrut \\
\hline
1 & $-0.026$ / $-0.002 \pm 0.010$ & $\phantom{-}0.000 \pm 0.007$ & $-0.002 \pm 0.012$ \Tstrut \\
2 & $\phantom{-}0.013$ / $\phantom{-}0.015 \pm 0.010$ & $\phantom{-}0.028 \pm 0.006$ & $-0.014 \pm 0.011$ \\
3 & $-0.001$ / $\phantom{-}0.006 \pm 0.010$ & $\phantom{-}0.021 \pm 0.006$ & $-0.014 \pm 0.012$ \\
4 & $\phantom{-}0.008$ / $\phantom{-}0.005 \pm 0.010$ & $\phantom{-}0.035 \pm 0.010$ & $-0.030 \pm 0.014$ \\
5 & $-0.011$ / $-0.005 \pm 0.010$ & $-0.018 \pm 0.028$ & $\phantom{-}0.013 \pm 0.030$ \\
6 & $-0.054$ / $-0.056 \pm 0.011$ & $-0.145 \pm 0.091$ & $\phantom{-}0.089 \pm 0.092$ \Bstrut \\
\hline
\end{tabular}   }
  \tablefoot{
    The values listed here are the fiducial mean and median SOM bias obtained from \skills\ (Sect.~\ref{SOM:fiducial}), followed by
    the shift-fit parameter obtained by fitting the CCs with the SOM redshifts, and finally the difference between the median SOM
    bias and the shift-fit parameter.
    The uncertainties of $\delta z_{\rm mean}$ and $\delta z_{\rm med}$ are identical to the third decimal place after applying
    the error floor. \dshift{SOM} is duplicated from Table~\ref{tab:CCfits} for comparison.
  }
\end{table}

This comparison also highlights that \dshift{SOM} is most closely comparable to the median SOM bias instead of the mean.
In particular, the mean redshift of bin one is very sensitive to small fractions of high-redshift, catastrophic outlier
populations, to which the measured CCs and the core of the \somnz, and in turn \dshift{SOM}, are insensitive.
 
\section{Discussion} \label{sec:discussion}
The results presented in the previous sections form the basis for measurements of WL with the KiDS-Legacy dataset.
Using extensive mock catalogues, we quantified the precision and accuracy of the redshift calibration of the six
tomographic bins used in those measurements. In the following, we highlight the most important aspects and lessons
learned from this calibration effort.

We relied heavily on mock catalogues that resemble the KiDS and KiDZ data in many important aspects (colour-redshift
relation, photometric noise level, photo-$z$ quality, clustering properties, etc.). This reliance makes it necessary to
introduce redundancy in the underlying simulations to test the robustness of the results to the assumptions in the
creation of the simulations. The two simulations used in this work are quite different. \skills\ is based on a
semi-analytic galaxy model, a full simulation of KiDS/VIKING images, and a replication of the KiDS photometry and
shape-measurement pipelines on these synthetic images. It also extends to high redshifts ($z<2.3$) covering the whole
redshift range of interest for KiDS. Our MICE2 mocks are more simplistic as we do not implement a full image simulation
here, employ a parametric model for photometric noise, and add shape-measurement weights in a rather ad hoc way. Also,
MICE2 is limited to $z<1.4$, which compromises its ability to calibrate the highest redshifts probed by KiDS. However,
it covers a much larger area than \skills, which makes it useful for clustering redshift analyses.

Most importantly, the two simulations are inherently so different that any agreement of the calibration results between
\skills\ (truncated at $z<1.4$ for comparison) and MICE2 can be regarded as a strong sign of systematic robustness. This
is exactly what we observe with the SOM calibration for the calibration samples that are most comparable between the two
simulations, meaning the samples without \lensfit\ weights (as those weights are not entirely realistic in MICE2); see
the $\Delta \langle\delta_z\rangle$ rows of scenarios $\left[{\textbf J}\right]$ (vs $\left[{\textbf H}\right]$) and
$\left[{\textbf K}\right]$ (vs $\left[{\textbf I}\right]$) in Table~\ref{tab:SOMbiases}. While the actual calibration
sample used in the cosmological analysis looks slightly different (i.e. using shape weights) and the redshift bias
values themselves will be different (e.g. compare $\langle \delta_z\rangle$ of scenarios $\left[{\textbf A}\right]$ and
$\left[{\textbf E}\right]$), there is no reason to assume that the robustness is affected by these differences. Hence,
we use the reported $\Delta \langle\delta_z\rangle$ to motivate a conservative systematic error floor of $\delta_z=0.01$
for our SOM results. Hence, the \skills\ SOM results, verified with the MICE2 SOM runs, yield a {primary},
simulation-based validation of the SOM \nz\ on the KiDS data at the per cent level in terms of the mean redshift, which
is exactly what is required for the full, uncompromised cosmological exploitation of cosmic shear with KiDS-Legacy.

The CC methodology with the KiDS-Legacy calibration samples is tested extensively on MICE2 and shown to be unbiased
within errors if the true \nz\ are used as a model in the shift fit. The goodness of fit is satisfactory when we inflate
the errors of the CC measurements and the noise according to what we observe on the data. The necessity for this
adaptation highlights a possible shortcoming of the MICE2 simulations that do not seem to replicate the full complexity
of systematic effects in the data. Still, with this adaptation, the similarity between the CC measurements on the data
and simulations gives us confidence in the applicability of the MICE2 results.

Most importantly for our efforts, the CC method is able to correct the bias inherent to the SOM \nz\ on MICE2 when those
are used as a model in the shift fit. This non-trivial result reported in Fig.~\ref{fig:fit_shifts} and
Table~\ref{tab:CCshifts} establishes the CC as a {secondary}, data-based method for \nz\ validation.

The crucial question is then whether the highly complementary {primary} and {secondary} validation methods agree on the
data. This is answered positively by Fig.~\ref{fig:fit_shifts_data}. The residual bias of the KiDS SOM \nz\ suggested by
\skills\ agrees with the bias suggested by shift-fitting the KiDS SOM \nz\ to the KiDS CC measurements. The only
exception is bin four, which exhibits a $2\sigma$ shift towards negative values; however, such a single shift will not
have a significant impact on the cosmological results.  The two methods of validation mentioned are similarly precise in
the first three tomographic bins. In bins five and six, the clustering redshifts still suffer from limited calibration
samples and possibly further systematics that affect the increasingly faint target samples, for example, spurious
density variations due to variable depth and seeing.

The residual biases and their uncertainties can be directly translated into priors on the mean redshifts used in the
cosmological inference of KiDS cosmic shear measurements. The discussion above motivates at least two main scenarios,
one that relies fully on the SOM \nz\ and their calibration with \skills\, and a second one using the SOM \nz\ but
calibrated with the CC measurements instead. It is clear that in the latter case the very loose priors on the mean
redshifts of bins five and six would severely compromise the constraining power of these bins. So this CC-calibrated
setup would constitute a very conservative approach. Even the tighter, \skills\--calibrated priors should still be
regarded as conservative because the error floor introduced due to residual differences between the SOM runs on \skills\
(truncated) and MICE2 is erring on the side of caution. There are very good reasons to believe in the superiority of the
\skills\ results. If we took those at face value, we would end up with priors on the mean redshifts that approach the
level of completed stage-IV cosmic shear surveys (see row $\sigma_{\delta z}$ of scenario $\left[{\textbf A}\right]$ of
Table~\ref{tab:SOMbiases}).
 
\section{Summary}\label{sec:summary}
In this paper we have presented the redshift calibration of the final KiDS WL dataset, dubbed KiDS-Legacy and based on
KiDS-\drfive. We developed a calibration strategy that involves multiple levels of redundancy to ensure that we met the
requirement of an accuracy of the mean redshifts of the tomographic bins used for cosmic shear at the percent level.

The first level of redundancy is represented by the use of two complementary sets of mock catalogues extracted from two
quite different types of simulations, \skills\ and MICE2. Using a newly developed matching algorithm, we arrive at mock
catalogues that are highly realistic and emulate the data (i.e. the KiDS WL sources as well as spectroscopic calibration
samples) with high fidelity.

The second level of redundancy is represented by two different calibration techniques, a colour-based SOM calibration
and a position based clustering redshift technique. This combination has become the standard in contemporary WL analyses
and is further strengthened here by an extensive overlap of KiDS-\drfive\ with different spectroscopic surveys and an
almost complete disconnect of the spectroscopic calibration samples used for either technique.

We show, in essence, that running the SOM on one simulation can be used to calibrate the WL sources in the other
simulation with residual bias $\langle\delta_z\rangle\la0.01$. Given that we estimate the match between the more
sophisticated simulation, \skills, and the KiDS data to be at least as good as the match between \skills\ and MICE2, we
are confident that \skills\ can calibrate the SOM \nz\ of KiDS at the same level of accuracy or better. The great
similarity between the \nz\ of the simulated \skills\ sources and the \nz\ estimated with the SOM on the data further
justifies the applicability of this conclusion to the KiDS dataset.

Additionally, we show that the clustering redshifts are able to correct for any residual bias in the SOM \nz\ on the
MICE2 simulation. With this result in mind, we ran the clustering redshift technique on the data, shift-fitting the SOM
\nz\ to the clustering measurements and found biases that agree with the purely simulated bias estimates from \skills.
This again mirrors the results of MICE2 clustering-$z$ vs \skills\ SOM \nz, which further validates the realism of the
simulations. Passing this strong consistency test suggests a robust calibration and a successful understanding and
correction of systematic errors at the percent level in terms of the mean redshifts of the tomographic bins.

These results have been used to define the priors on the mean redshifts of the tomographic bins in the
cosmic shear analyses of KiDS-Legacy \citep{wright/etal:2025b,stoelzner/etal:2025}. 
A \skills\--based SOM \nz\ calibration with a conservative error floor
constitutes the fiducial setup, taking the form of informative correlated Gaussian priors with means $\delta z = \{ -2.6
, 1.4 , -0.2 , 0.8 , -1.1 , -5.4 \}\times 10^{-2}$ and standard deviation $\sigma_{\delta z}\approx0.01$.  As an
alternative, we also present a purely empirical, somewhat less constraining setup that takes the clustering
redshift results as (differently correlated) priors that make the cosmological conclusions independent of simulations of
the redshift calibration.

With the type of data used here, we reach statistical uncertainties on the mean redshifts with our SOM implementation of
$\sigma(\langle\delta_z\rangle)\approx0.002$, which falls within the range of the requirement for \euclid. This suggests
that, in terms of methodology and calibration data, we are almost ready to calibrate a stage-IV cosmic shear survey.
Certainly, the redshift range has to be extended to $z\sim2$, but this is well within reach. The real challenge will be
to reduce the systematic error floor, conservatively estimated here, by about a factor of 5. This will require a set of
a few highly realistic, complementary simulations that capture the whole complexity of a future cosmic shear experiment.

The statistical uncertainties on the clustering redshifts shown here are still at least a factor of 3 greater than those
required by \euclid.  With an order of magnitude more area in the WL samples and the full power of upcoming wide-field
spectroscopic calibration samples (a glimpse is given here with just $\sim10^5$ DESI EDR galaxies), this factor of 3 is
within reach. Systematic error control will be paramount here as well, and will be similarly achieved through redundancy
in the simulations that validate the calibration.
 
\begin{acknowledgements}
AHW and HHi are supported by the Deutsches Zentrum für Luft- und Raumfahrt (DLR), made possible by the Bundesministerium
  für Wirtschaft und Klimaschutz, under project 50QE2305, and acknowledge funding from the German Science Foundation DFG, via the Collaborative Research Center SFB1491 ``Cosmic Interacting Matters - From Source to Signal''.
HHi is also supported by a DFG Heisenberg grant (Hi 1495/5-1). HHi, JLvdB, CM, RR, \& AD are supported by an ERC Consolidator Grant (No. 770935).
MB \& PJ are supported by the Polish National Science Center through grant no. 2020/38/E/ST9/00395. MB is also supported by grant no. 2020/39/B/ST9/03494.
CH, BS, \& ZY acknowledge support from the Max Planck Society and the Alexander von Humboldt Foundation in the framework of the Max Planck-Humboldt Research Award endowed by the Federal Ministry of Education and Research. CH also acknowledges support from the UK Science and Technology Facilities Council (STFC) under grant ST/V000594/1.
BJ acknowledges support by the ERC-selected UKRI Frontier Research Grant EP/Y03015X/1 and by STFC Consolidated Grant ST/V000780/1.
CM acknowledges support from the Beecroft Trust, and the Spanish Ministry of Science under the grant number PID2021-128338NB-I00.
MA is supported by the UK Science and Technology Facilities Council (STFC) under grant number ST/Y002652/1 and the Royal Society under grant numbers RGSR2222268 and ICAR1231094. CG is funded by the MICINN project PID2022-141079NB-C32.
BG acknowledges support from the UKRI Stephen Hawking Fellowship (grant reference EP/Y017137/1).
HHo \& MY acknowledge support from the European Research Council (ERC) under the European Union’s Horizon 2020 research and innovation program with Grant agreement No. 101053992.
SJ acknowledges the Ram\'on y Cajal Fellowship (RYC2022-036431-I) from the Spanish Ministry of Science and the Dennis Sciama Fellowship at the University of Portsmouth. 
KK acknowledges support from the Royal Society and Imperial College.
SSL is receiving funding from the programme ``Netzwerke 2021'', an initiative of the Ministry of Culture and Science of the State of North-Rhine Westphalia.
LL is supported by the Austrian Science Fund (FWF) [ESP 357-N].
AL acknowledges support from the research project grant `Understanding the Dynamic Universe' funded by the Knut and Alice Wallenberg Foundation under Dnr KAW 2018.0067.
LM acknowledges the financial contribution from the grant PRIN-MUR 2022 20227RNLY3 “The concordance cosmological model: stress-tests with galaxy clusters” supported by Next Generation EU and from the grant ASI n. 2024-10-HH.0 `Attività scientifiche per la missione Euclid – fase E'.
LP acknowledges support from the DLR grant 50QE2002.
MR acknowledges financial support from the INAF grant 2022.
TT acknowledges funding from the Swiss National Science Foundation under the Ambizione project PZ00P2\_193352.
MvWK acknowledges the support by the UKSA and STFC (grant no. ST/X001075/1).
YZ acknowledges the studentship from the UK Science and Technology Facilities Council (STFC).
Based on observations made with ESO Telescopes at the La Silla 
Paranal Observatory under programme IDs 179.A-2004, 177.A-3016, 177.A-3017, 
177.A-3018, 298.A-5015. 
\\
\textit{Author Contributions:} 
All authors contributed to the development and writing of this paper. The
authorship list is given in three groups: the lead authors (AHW, HHi, JLvdB), 
followed by two alphabetical groups. The first alphabetical
group includes those who are key contributors to both the scientific analysis
and the data products of this manuscript and release. The second group covers those who 
have made a significant contribution either to the preparation of data products or to the 
scientific analyses of \kidslegacy. 
\end{acknowledgements}

\bibliographystyle{aa}
\bibliography{library}

\appendix 

\section{Sample selection function for MICE2}\label{app:mocks}

In our clustering redshift analysis with simulated data based on MICE2 we aimed to replicate the observational datasets
as closely as possible. Since we added spectroscopic data from the DESI EDR and VIPERS \mbox{PDR-2}, we also needed to
implement their selection functions for MICE2, similar to the procedure already adopted in
\citet{vandenbusch/etal:2020}. Where possible, we applied the same selection criteria used for the spectroscopic target
selection and used sampling strategies to implement additional selection effects, such as spectroscopic success rates,
and to mitigate systematic differences between MICE2 and the observed datasets.

\subsection{DESI EDR data}
We used a subset (called the LSS catalogues) of the DESI LRG and ELG samples for the \kidslegacy\ clustering redshifts.
The target selection for the ELG sample is a simple colour-magnitude cut, which can, in principle, be applied directly
to MICE2. The LRG sample, however, and some of the additional selections applied for the construction of the LSS
catalogues, depend on observed quantities to which we do not have access to in MICE2. Therefore, we decided to implement
the DESI selection function for MICE2 by mostly relying on sampling techniques.

\begin{figure}
  \centering
  \includegraphics[width=\columnwidth]{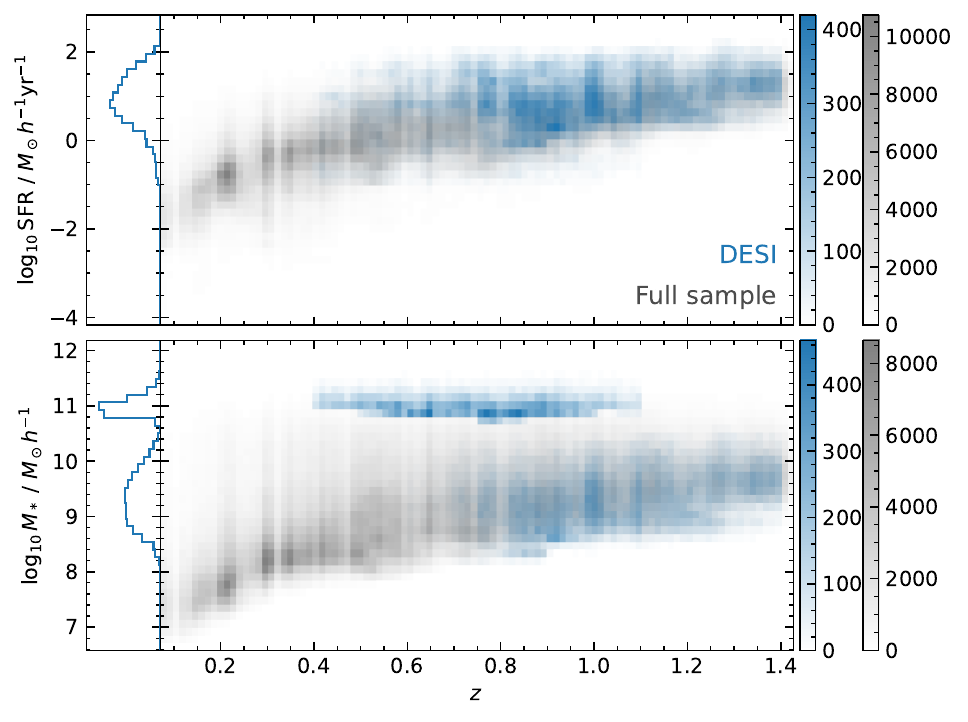}
  \caption{Comparison of the star-formation rate (top) and the stellar mass (bottom) for the full MICE simulation (grey)
  and our simulated DESI LRG and ELG samples (blue) as a function of redshift. The data is selected from a $44~\sqdeg$
  patch of MICE2. The lower panel clearly shows the separation of the ELG from the LRG sample, which is selected based
  on stellar mass.
  }
  \label{fig:desi_selection}
\end{figure}

We selected the LRG and ELG sample jointly by performing a number of selection steps. First, we split the DESI data and
MICE2 into bins of redshift ($\Delta z = 0.05$). In each of these bins we computed the expected number of ELGs, LRGs, and
(although not utilised) quasars (QSOs). Then we randomly draw the appropriate number QSOs from MICE2 by
requiring $19.5 < r < 23.4$ without any further selections (MICE2 does not contain any QSOs specifically) and discarded 
them. From the remaining MICE2 galaxies we then selected the expected number of ELGs by picking the objects with the
highest specific star formation rate that fell in the magnitude window $20.0 < g < 24.1$. Finally, we drew the
expected number of LRGs from those MICE2 galaxies that were not already assigned to either the QSO or ELG sample. We
selected objects with the highest stellar mass, and at magnitude $z < 21.61$. This procedure ensures that the MICE2 DESI
sample has the correct redshift distribution by design. Figure~\ref{fig:desi_selection} shows the distribution of
stellar mass and star-formation rate of galaxies in all of MICE2 and those in our simulated DESI subset.  The LRG and
ELG subsets are clearly separated in stellar mass, the ELG sample contains mostly objects with low stellar mass but high
star-formation rate.

\begin{figure}
  \centering
  \includegraphics[width=\columnwidth]{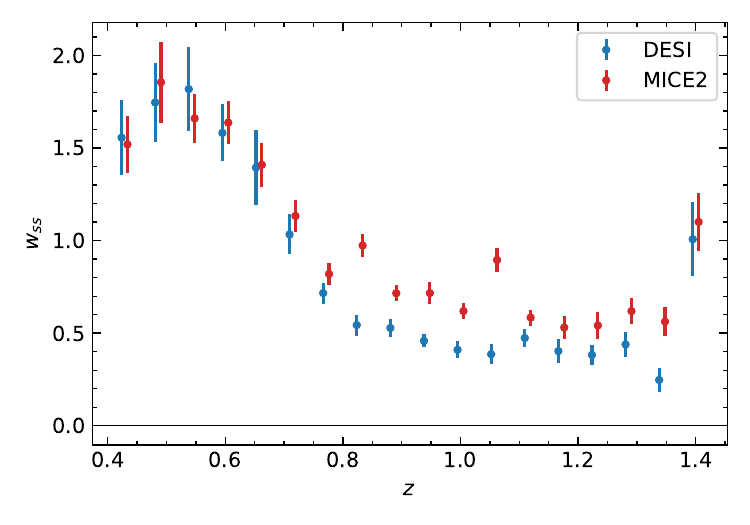}
  \caption{Auto-correlation amplitude measured between 100 and 1000$\,$\kpc{} for DESI (in blue) and MICE2 (in red).
  }
  \label{fig:desi_wss}
\end{figure}

To verify our new selection function we compared its clustering amplitude $w_{\rm ss}$ with the one we obtain from DESI. We
measured the angular correlation between 100 and 1000$\,$\kpc{} and found good agreement between simulation and data for most
redshifts, except around the redshifts $0.85$ and $1$ (see Fig.~\ref{fig:desi_wss}). Since we were already
selecting objects with low stellar mass in our ELG selection (for which we already expect a lower galaxy bias and therefore a
lower clustering amplitude; see e.g. \citealt{coil/etal:2017}), we speculate that this may be an inherent property of MICE2.

\subsection{VIPERS data}

VIPERS targets galaxies with magnitude $i_{\rm AB} \le 22.5$ and an additional colour selection that aims to isolate
galaxies at $z > 0.5$ \citep{scodeggio/etal:2018}:
\begin{equation}
  (r - i) > 0.5 \times (u - g) \quad\text{OR}\quad (r - i) > 0.7 \;.
\end{equation}
We applied the same selection criteria to MICE2.

\begin{figure}
  \centering
  \includegraphics[width=\columnwidth]{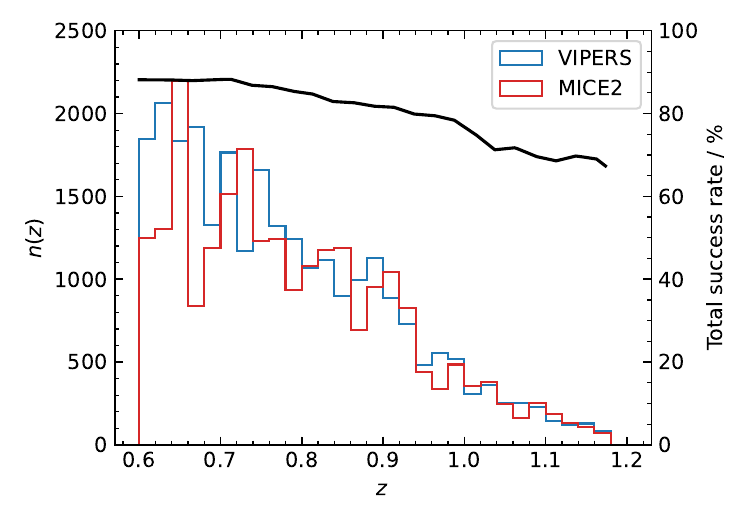}
  \caption{Comparison of the VIPERS redshifts distribution (in blue) in the range $0.6 \le z < 1.18$ and MICE2 (in red)
  after applying the colour and magnitude cuts, and the
  empirical incompleteness sampling, indicated by the total success rate (black line).
  }
  \label{fig:vip_nz}
\end{figure}

This colour selection (colour sampling rate) leads to a completeness that transitions from almost zero to one in the
range of $0.4 < z < 0.6$. There are two additional effects that need to be factored in to obtain the total completeness
of the sample; the target sampling rate (TSR), which is about 50~\% on average but has a strong positional dependence
due to observational and instrumental limitations, and the spectroscopic success rate (SSR). The total completeness
is a product of these three terms and VIPERS defines a weight to account for this incompleteness as
\begin{equation}
  w = \frac{1}{{\rm CRS} \times {\rm TSR} \times {\rm SSR}} \;.
\end{equation}
For our purposes, we chose to not model the positional dependence of the TSR and simply estimate the mean incompleteness
weight empirically in the redshift range $0.6 \le z < 1.18$ (see Sect.~\ref{sec:calib_data}), as shown in
Fig.~\ref{fig:vip_nz}. When applied to MICE2 together with the VIPERS colour selection, we found that this approach
reproduced the redshift distribution $p(z)$ of the VIPERS dataset very well. However, we needed to apply an additional
sparse sampling by 30~\% to match the absolute number density found in the data. Similar discrepancies have been
reported by \citet{vandenbusch/etal:2020} when trying to reproduce the selection functions of other high redshift
datasets and are most likely explained by systematic differences between MICE2 and the observational data.

\begin{figure}
  \centering
  \includegraphics[width=\columnwidth]{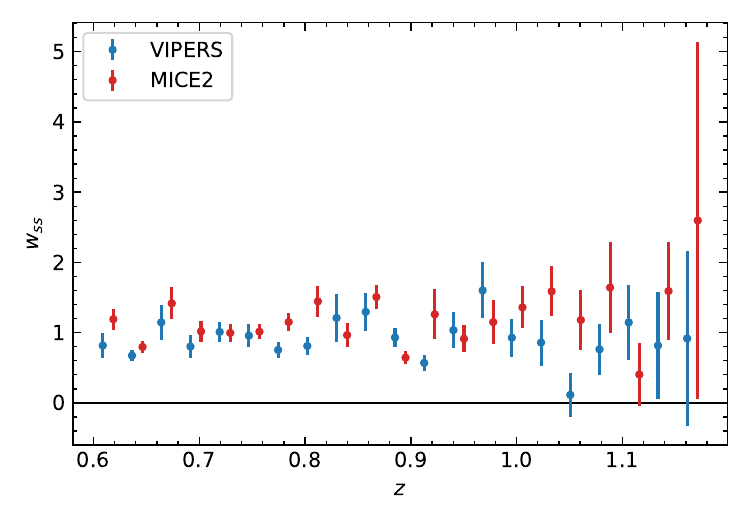}
  \caption{Auto-correlation amplitude measured between 100 and 1000$\,$\kpc{} for VIPERS (in blue) and MICE2 (in red).
  }
  \label{fig:vip_wss}
\end{figure}

Similar to DESI, we verified our new selection function by comparing its clustering amplitude $w_{\rm ss}$ with the one we
obtained from VIPERS. We measured on the same scales from 100 and 1000$\,$\kpc{} and found a good agreement between
simulation and data, both in the amplitude as well as its uncertainty over the full redshift baseline
(Fig.~\ref{fig:vip_wss}).

\end{document}